\documentclass[aps,prx,twocolumn,showpacs,amsmath,amssymb]{revtex4}
\usepackage{epsfig}
\usepackage{graphicx}
\usepackage{bm}
\usepackage{braket}
\usepackage[dvips]{color}
\usepackage{mathrsfs}
\usepackage{sidecap}
\newcommand{\aup}{\hat{a}^{\dag}}
\newcommand{\adown}{\hat{a}}
\newcommand{\cp}{\hat{c}^{\dag}}
\newcommand{\cdown}{\hat{c}}
\newcommand{\dup}{\hat{d}^{\dag}}
\newcommand{\ddown}{\hat{d}}

\begin{document}
\title{Non-local transport properties of nanoscale conductor-microwave cavity systems}

\author{C. Bergenfeldt and P. Samuelsson} \affiliation{Division of
Mathematical Physics, Lund University, Box 118, S-221 00 Lund, Sweden}

\begin{abstract} 
Recent experimental progress in coupling nanoscale conductors to
superconducting microwave cavities has opened up for transport
investigations of the deep quantum limit of light-matter interactions, with
tunneling electrons strongly coupled to individual cavity photons. We have
investigated theoretically the most basic cavity-conductor system with strong, single photon induced
   non-local transport effects; two spatially separated double quantum
   dots (DQD:s) resonantly coupled to the fundamental cavity mode. The system, described by a generalized Tavis-Cummings
   model, is investigated within a quantum master equation formalism,
   allowing us to account for both the electronic transport properties
   through the DQD:s as well as the coherent, non-equilibrium cavity
   photon state. We find sizeable non-locally induced current and current
   cross-correlations mediated by individual photons. From a full statistical description of the electron transport
   we further reveal a dynamical channel blockade in one DQD lifted by
   photon emission due to tunneling through the other DQD. Moreover,
   large entanglement between the orbital states of electrons in the
   two DQD:s is found for small DQD-lead temperatures.

\end{abstract}

\pacs{xx}
\maketitle 

\section{Introduction}

In circuit quantum electrodynamics (QED), the mesoscopic analog of cavity
QED, solid state qubits are coupled to
superconducting microwave cavities on chip \cite{Blais04,Wall2004}. 
Circuit QED systems combine the appealing properties of high cavity quality
factors and strong vacuum microwave fields with low
qubit decoherence. This has allowed for experiments in the strong coupling
limit with qubit-cavity coupling exceeding the qubits decoherence rates. The
strong cavity-qubit coupling together with fast, coherent
manipulation of the qubits has lead to an astonishing development in
the areas of quantum information processing \cite{Schoel2008,Sillanpaa07,Majer07,DiCarlo09,DiCarlo10} and microwave quantum
optics with superconducting circuits
\cite{hofheinz08,hofheinz09,Eichler10,Eichler11,Wang11,Houck07}. Moreover,
circuit QED architectures have a large potential for simulations of strongly
interacting many-body systems \cite{Lehur} and tests
of fundamental quantum physical effects \cite{Johansson09,Wilson11,Vijay11}.

The rapid development in circuit QED triggered investigations on nanoscale qubits
and conductors coupled to microwave cavities or
resonators \cite{Childress04,Burkard06,Astafaiev07,Trif08,Guo08,Marthaler08,Lambert09,Cottet102,Basset10,hofheinz11,Frey11,Delbecq11,Cottet11,Jin11,Bergenfeldt12,Cottet12,Jin12,Frey212,Chernii12,Frey112,Toida12,Petersson12,Delbecq13}.
Particularly interesting are recent experiments on few-level quantum dots
coherently coupled to microwave cavities \cite{Frey11,Delbecq11,Frey112,Toida12,Petersson12,Frey212,Delbecq13}. These
experiments open up for transport investigations of light-matter
interactions in the deep quantum regime: single electrons interacting
strongly and coherently with individual microwave photons. The large
versatility of microwave photon state properties \cite{Sandberg08,Wilson10,Wilson11},
together with the well-established controllability of quantum dot levels provide a broad scope for fundamentally
important experiments.

\begin{figure}[ht]
\begin{center}
\includegraphics[trim=0cm 0cm 0cm 0cm, clip=true, width=0.45\textwidth, angle=0]{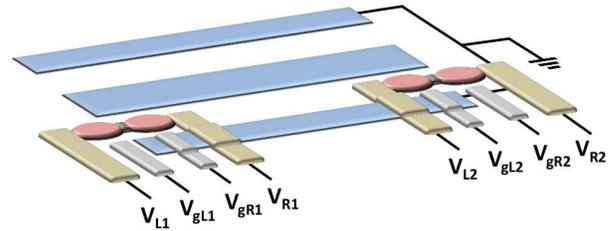}
\end{center}
\caption{Two DQD-cavity system: Each DQD (pair of red ovals) is coupled to the central
  conductor of the transmission line
  cavity (middle blue rectangle), two lead electrodes (gold rectangles) and to
  two gate electrode potentials (silver rectangles). Here $V_{L(R)i}$ and $V_{gL(R)i}$ denote the left (right) lead and gate electrode potential of DQDi, respectively.\label{newsys}}
\end{figure}

A key feature of 
conductor-cavity systems is the possibility to coherently couple electrons
in  conductors separated up to centimeters \cite{Delbecq13}. This puts
in prospect entangling macroscopically separated transport electrons, of
importance for nanoscale quantum information processing
and Bell inequality tests
\cite{Chtchelkatchev02,Samuelsson03,Beenakker03,Samuelsson04}. Moreover,
this non-local feature can be harnessed for efficient
heat-transfer or refrigeration over large distances \cite{Meschke06,Timofeev09}.   
A first step towards these goals would be an experimental demonstration of
non-local, few-photon mediated, electronic transport effects. In the present work we investigate theoretically the simplest possible
strongly coupled cavity-quantum dot system where such non-local
effects can be observed: two double quantum dots (DQD:s)
coupled to the same transmission line cavity (See Fig. \ref{newsys}). We argue
that dot-cavity systems with single-level or metallic dots will, in comparison, display strongly surpressed non-local effects.

The DQD:s are resonantly coupled to the same, fundamental mode of the microwave cavity.
This DQD-cavity system constitutes a generalized Tavis-Cummings model
\cite{Tavis68,Fink09}, with strong hybridization of the DQD electron and microwave
photon states. Our investigation is focused
on the non-local electronic transport properties. For a broad range of
parameters, the current-voltage characteristics provide clear signatures of
transport electrons exchanging photons. Further,
for asymmetric DQD-lead couplings, noise and higher order current fluctuations
reveal a dynamical channel blockade in one DQD lifted by 
single photons emitted by electron-tunneling in the second DQD. In addition, we
demonstrate the existence of large orbital entanglement between
electrons in different DQD:s.

We emphasize that the predicted non-local effects are direct consequences 
of the non-equilibrated, transport-induced photon state.
This makes our investigation qualitatively different from earlier transport studies
on pairs of two-level systems coupled via thermalized bosons \cite{Vorrath03,Contreras08}.
To fully account for the coherent, non-equilibrium properties of the photon state, as well as the electron tunneling through the DQD:s, our investigation is carried out within the
framework of a quantum master equation (QME). The approach is
  similar to the ones used to investigate transport through single two-level systems coupled resonantly to a photonic mode in Refs. \cite{Marthaler08,Lambert09,Jin12}.

The article is organized as follows: In Sec. \ref{model} we introduce the model
for the closed DQD-cavity system, present its Hamiltonian and discuss the eigenstates and
eigenenergies. We further derive, in Sec. \ref{derQME}, a QME for the system when
the DQD:s are coupled to lead electrodes. In Sec. \ref{nonLocal} the
transport properties are investigated in the regime where the temperature of
the lead electrodes exceeds the DQD-cavity coupling strength. Focus is put on the
non-local transport properties, calculating the non-local current-voltage characteristics,
the current cross-correlations and the full counting statistics. 
In Sec.  \ref{TavisCummingsfine} we turn to the regime where the lead
temperature is smaller than the DQD-cavity coupling strength. 
Transport signatures of coherent electron-photon interaction as well as
entanglement between the electrons in the DQD:s are investigated. The
effect of dephasing and approaches to minimize this effect are discussed in Sec. \ref{Exp}.

\section{System and method}
\label{sys}
The system considered is depicted in Fig. \ref{newsys}. Two DQD:s, denoted 1
and 2, are inserted near the endpoints of a
transmission line cavity. The central conductor is
capacitively coupled to the right (left) dot in DQD1(2) (see e.g.
Ref. \cite{Frey112} for a possible experimental realization). 
One gate and one lead electrode is further coupled to each dot
in the DQD:s. The leads are assumed to be in 
thermal equilibrium with a common temperature $T$ and chemical potentials
$\mu_{\nu i}$, with $\nu=L,R$ and $i=1,2$ denoting to which dot the lead is
coupled.

Throughout most of the paper we will consider the strong coupling limit. This
implies that the DQD-cavity coupling is large compared to the DQD-lead couplings
and also dominates over decoherence due to other type of system-environment
interaction. Moreover, we will the DQD-lead couplings to be much stronger than the
interaction with the rest of the system-environment and hence neglect decoherence from the latter. Only in last section the effect of DQD dephasing as well as DQD-relaxation and photon loss will be considered.

\subsection{Model}
\label{model}
The DQD:s, forming singly occupied two-level systems, couple linearly to the
microwave photons in the cavity. 
The system Hamiltonian $\hat{H}_{S}$, describing the DQD-photon interaction as
well as the orbital degrees freedom of the DQD:s and the direct interaction between the DQD
charges, is derived in Appendix A. Below we will take the DQD:s to be on
resonance with the fundamental mode of the transmission line cavity. Moreover, the cavity characteristic
impedance $Z_{0}$ is assumed to be much smaller than the resistance
quantum $R_{Q}=h/e^{2}$, relevant for regular transmissionline cavities. Under
these conditions the DQD-cavity system will be described by a generalized Tavis-Cummings (TC) Hamiltonian \cite{Tavis68}
\begin{align}
&\hat{H}_{S}=\hbar\omega\aup\adown+\sum_{i}\bigg[\frac{\hbar\omega}{2}(\hat{d}^{\dag}_{e
  i}\hat{d}_{ei}-\hat{d}^{\dag}_{gi}\hat{d}_{gi})\nonumber\\
&+\hbar g_{0}(\aup\dup_{gi}\ddown_{ei}+h.c.)\bigg].
\label{Tavis}
\end{align}
Here $\dup_{gi}$ and $\dup_{ei}$ ($\ddown_{gi}$ and $\ddown_{ei}$) denotes the
creation (annihilation) operators of
the ground and exited, i.e. of the bonding and anti-bonding, state of DQDi. 
We have further introduced the photon creation operator $\aup$
and the frequency $\omega$ of the fundamental mode, 
and the DQD-cavity coupling strength $g_{0}$, for simplicity taken equal for both DQD:s.
Note that since we have assumed single occupancy of the DQD:s, the spin-degree
of freedom of the DQD:s will only have the effect of renormalizing tunneling
rates and is hence neglected in Eq. \eqref{Tavis} and below.

The generalized TC Hamiltonian in Eq. \eqref{Tavis} has the form of a
TC Hamiltonian for both DQD:s occupied while it reduces to a
Jaynes-Cummings (JC) and a harmonic oscillator (HO) Hamiltonian when one or
none of the DQD:s are occupied, respectively.
It is of key importance for the discussion below to describe the eigenstates in
the HO, JC and TC subspaces of $\hat{H}_{S}$. We first note that $\hat{H}_{S}$
commutes with the operator for the number of excitations
$\hat{n}=\aup\adown+\sum_{i}\dup_{ei}\ddown_{ei}$.
The eigenstates can then be characterized by the corresponding quantum number $n$. 
We express the eigenstates in terms of the DQD-cavity product states  
$\ket{\xi_{1}\xi_{2} p}$, with DQDi in the state $\ket{\xi_{i}}$, with $\xi
_{i}=0,g,e$, and $p$ photons in the cavity mode.

For the HO subspace
$\xi_{1}=\xi_{2}=0$ and hence the number of excitations is equal to the
number of photons, giving the eigenstates $\ket{00n}$. 
In the JC subspace with DQD1(2) occupied the eigenstate with zero excitations is
$\ket{S_{1}^{0}0}=\ket{g00}(\ket{S_{2}^{0}0}=\ket{0g0})$. For states with a finite number
of excitations the photon state and the state of the occupied DQD
hyridises. The states, denoted by $\ket{S_{i}^{\pm}n}$, are superpositions of
product states with $n$ and $n-1$ photons in the mode. For DQD1 
occupied they are given by $\ket{S_{1}^{\pm}n}=[\ket{g
  0n}\pm\ket{e0n-1}]/\sqrt{2}$ and for DQD2 occupied we have $\ket{S_{2}^{\pm}n}=[\ket{
0gn}\pm\ket{0en-1}]/\sqrt{2}$. The eigenbasis in the
TC subspace has a similar structure. The state with zero excitations is a
product state $\ket{D^{0}0}=\ket{gg0}$ and the states with one or more excitations
 are superpositions of product states with different number of photons.
We denote the finite-excitation eigenstates by
$\ket{D^{\alpha}1}$, with $\alpha=0,\pm$, for $n=1$ and 
$\ket{D^{\beta\gamma} n}$, with
 and $\beta,\gamma=\pm$, for $n\geq 2$ and give their exact forms in Appendix B.  
The spectra of the HO, JC and TC Hamiltonians, also given in Appendix B, are shown in
Fig. \ref{energydiag}. Importantly we see in Fig. \ref{energydiag} that a state with $n$
excitations has an energy $n\hbar\omega+\mathcal{O}(\hbar g_{0})$ relative to
the energy of the state with zero excitations in its subspace. Moreover the
TC (HO) ground state is shifted $-\hbar\omega/2$ ($\hbar\omega/2$) with respect to
the JC ground states. 
\begin{figure}[ht]
\begin{center}
\includegraphics[trim=0cm 0cm 0cm 0cm, clip=true, width=0.50\textwidth, angle=0]{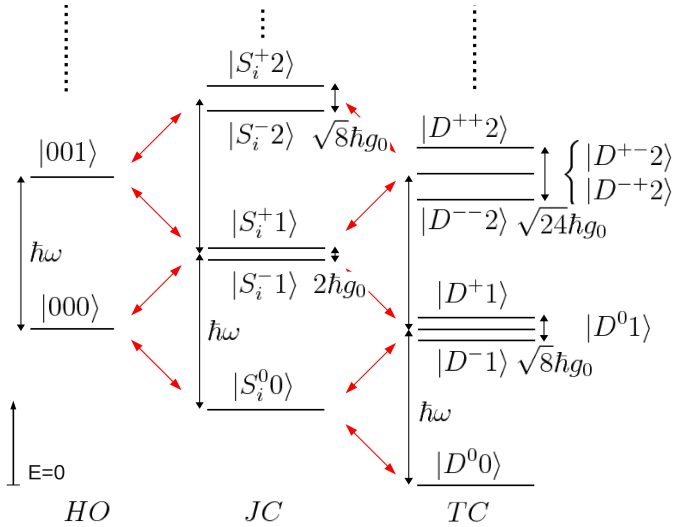}

\caption{The spectrum of the Hamiltonian in Eq. \eqref{Tavis} with the
  eigenenergies marked by the horizontal black lines. Column to the left
  shows the spectrum of the HO subspace (both DQD:s unoccupied), 
middle column shows JC spectrum (DQD1 or DQD2 occupied)
    and right column shows the spectrum of the TC subspace (both DQD:s
    occupied). The diagonal red arrows show the allowed transitions due to
    electron tunneling described by Eq. \eqref{QME}. \label{energydiag}\label{Tavisspec}
}
\end{center}
\end{figure} 

\subsection{Quantum Master Equation}
\label{derQME}
The leads and their tunnel coupling to the DQD:s are
described by the Hamiltonians $\hat{H}_{L}$ and $\hat{H}_{T}$, respectively.
The lead Hamiltonian reads $\hat{H}_{L}=\sum_{k,\nu,i}\epsilon_{k}\cp_{k\nu
  i}\cdown_{k\nu i}$, with $\cp_{k\nu i}$ creating an electron in
the state with energy $\epsilon_{k}$ in the lead connected to dot $\nu i$.
In the ground-excited basis the tunneling Hamiltonian is given by
\begin{align}
&\hat{H}_{T}=\sum_{k,i}\Bigg(t_{Li}\cp_{kLi}\left[-\sin(\theta_{i})\hat{d}_{gi}+\cos(\theta_{i})\hat{d}_{ei}\right]\nonumber\\
&+t_{Ri}\cp_{kRi}\left[\cos(\theta_{i})\hat{d}_{gi}+\sin(\theta_{i})\hat{d}_{ei}\right]+h.c\Bigg),
\end{align}
with $t_{\nu i}$ denoting the energy independent lead-dot tunneling amplitude for dot
$\nu i$. We have further introduced the DQD mixing angles
$\tan(\theta_{i})=\pm t_{LRi}/(\sqrt{\omega^{2}-t_{LRi}^{2}}\pm\omega)$, where
$t_{LRi}$ denotes the interdot tunneling amplitude of DQDi. The +(-) sign
here refers to the energy difference between orbitals of the left and right dots of
DQDi being positive (negative).

We assume weak tunnel couplings between the dots and the leads
and restrict the investigation to the sequential tunneling regime. Following the standard Born-Markov approximation scheme a quantum
master equation (QME) is derived for the time evolution of the reduced density
matrix $\hat{\rho}$ of the DQD-cavity system \cite{qnoise}. We point out that the
lead-dot tunneling rates $\Gamma_{\nu i}=2\pi|t_{\nu i}|^{2}\sum_{k}\delta(\epsilon-\epsilon_{k})$ must be chosen much smaller the DQD-cavity coupling
strength $g_{0}$. This restriction is necessary for the strong coupling condition to
hold. Moreover, it allows us to neglect coherences between states with an energy difference
$\Delta\epsilon\gtrsim\hbar g_{0}$, i.e. to perform a secular approximation. Considering for simplicity identical
tunnel-couplings to the left and right dot in each DQD, i.e. $\Gamma_{Li}=\Gamma_{Ri}=\Gamma_{i}$, we can write the QME $\frac{d\hat{\rho}}{dt}=\mathcal{L}[\hat{\rho}]$, with
the Liouvillian
\begin{align}
&\mathcal{L}[\hat{\rho}]=-\frac{i}{\hbar}[\hat{H}_{S},\hat{\rho}]-\sum_{\nu,i}\sum_{\xi=e,g}\int{d\epsilon
d\epsilon'}\bar{\Gamma}_{\nu\xi i}(\theta_{i})\nonumber\\
\times\Bigg[&f_{\nu
    i}(\epsilon)\ddown_{\xi
    i}\delta(\epsilon+\epsilon'-\hat{H}_{S})\dup_{\xi i}\delta(\epsilon'-\hat{H}_{S})\hat{\rho}\nonumber\\
+&\tilde{f}_{\nu
    i}(\epsilon)\dup_{\xi
    i}\delta(\epsilon+\epsilon'+\hat{H}_{S})\ddown_{\xi i}\delta(\epsilon'+\hat{H}_{S})\hat{\rho}\nonumber\\
-&f_{\nu i}(\epsilon)\delta(\epsilon+\epsilon'+\hat{H}_{S})\dup_{\xi i}\delta(\epsilon'+\hat{H}_{S})\hat{\rho}\ddown_{\xi
  i}\nonumber\\
-&\tilde{f}_{\nu i}(\epsilon)\delta(\epsilon+\epsilon'-\hat{H}_{S})\ddown_{\xi i}\delta(\epsilon'-\hat{H}_{S})\hat{\rho}\dup_{\xi
  i}+h.c.\Bigg].
\label{QME}
\end{align}
Here $f_{\nu i}(\epsilon)=f(\epsilon-\mu_{\nu i})$ and $\tilde{f}_{\nu
  i}(\epsilon)=1-f(\epsilon-\mu_{\nu i})$, with $f$ denoting the Fermi-function. 
We have further introduced the rates
$\bar{\Gamma}_{Lei}(\theta_{i})=\bar{\Gamma}_{Rgi}(\theta_{i})=\Gamma_{i}\cos^{2}(\theta_{i})$ and
$\bar{\Gamma}_{Lgi}(\theta_{i})=\bar{\Gamma}_{Rei}(\theta_{i})=\Gamma_{i}\sin^{2}(\theta_{i})$.

\begin{figure}[ht]
\begin{center}
\includegraphics[trim=0cm 0cm 0cm 0cm, clip=true, width=0.45\textwidth, angle=0]{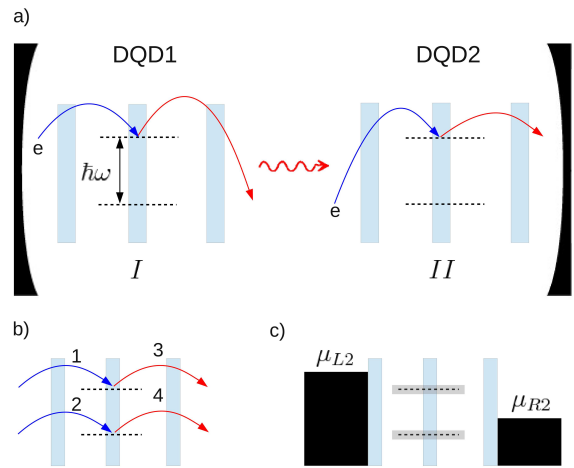}
\caption{(a) Elementary process in which two transport electrons exchanges a photon. In
  step I an electron tunnels through DQD1,
  i.e. into (blue arrow) and then out of (red arrow) DQD1, while increasing the
  system energy by $\hbar\omega$. In step II an electron tunnels through 
  DQD2 and changes the system energy by $-\hbar\omega$. Through steps I and II the two
  electrons have exchanged a photon (whiggling line) through the cavity. (b)
  Tunneling processes into and out the ground and excited state of DQDi decoupled from the cavity mode. The rates for the processes 1,2,3, and 4 are $\bar{\Gamma}_{L
    ei}f_{L i}(\hbar\omega/2)$, $\bar{\Gamma}_{Lgi}f_{L
   i}(-\hbar\omega/2)$, $\bar{\Gamma}_{R
    ei}\tilde{f}_{Ri}(\hbar\omega/2)$ and
  $\bar{\Gamma}_{Rgi}\tilde{f}_{Ri}(-\hbar\omega/2)$. (c) Bias configuration
  of DQD2 considered in Sec. \ref{nonLocal}. The positions of the chemical potentials of the
  left leads $\mu_{L2}$ and $\mu_{R2}$ relative to the energies 
  at which electron tunnel into and out of the DQD are shown. These sets of energies are in
  an interval $\sim\hbar g_{0}$ denoted by the grey boxes around the dashed lines.\label{exchange}}
\end{center}
\end{figure}

We point out that electron tunneling into or out
of the DQD:s in the sequential secular regime can be associated with jumps between
energy levels in adjacent columns in Fig. \ref{Tavisspec}. This means that the most compelling
physical picture will emerge when working in the
eigenbasis of $\hat{H}_{S}$ (see Sec. \ref {model}). By evaluating the rates
for the different tunneling processes, described by Eq. \eqref{QME}, in this
eigenbasis we can draw the following two important conclusions:

$(i)$ The number of excitations of the system can change with $0$ or $-(+)1$
when an electron tunnels into (out of) one of the DQD:s, as indicated by the red
arrows in Fig. \ref{Tavisspec}. The system energy can
thus increase or decrease by the energy $\hbar\omega/2+\mathcal{O}(\hbar g_{0})$
in a tunneling event. Consequently, an electron
tunneling through one of the DQD:s, e.g. from the left lead into the DQD and
then out to the right lead, can change the system energy by $0,\pm\hbar\omega$
$+\mathcal{O}(\hbar g_{0})$. If a tunneling electron changes the energy with
$+(-)\hbar\omega+\mathcal{O}(\hbar g_{0})$ a subsequently tunneling electron,
in the same or the other DQD, can absorb (emit) 
this energy. This process, here referred to as the \textit{transport electrons
exchanging a photon}, is depicted in Fig. \ref{exchange}.

 $(ii)$ The rate for processes where the energy is increased or decreased for an 
electron tunneling into dot $\nu$ in DQDi is proportional to 
 $\bar{\Gamma}_{\nu ei}f_{\nu i}[\hbar\omega/2+\mathcal{O}(\hbar g_{0})]$ and
 $\bar{\Gamma}_{\nu gi}f_{\nu
   i}[-\hbar\omega/2+\mathcal{O}(\hbar g_{0})]$, respectively. Similarly, the
 rates for exciting or
 deexciting the system when electrons tunnel out of DQDi are given by 
$\bar{\Gamma}_{\nu gi}\tilde{f}_{\nu i}[-\hbar\omega/2+\mathcal{O}(\hbar g_{0})]$ and
$\bar{\Gamma}_{\nu ei}\tilde{f}_{\nu i}[-\hbar\omega/2+\mathcal{O}(\hbar g_{0})]$.
Note that the rates for tunneling events in which energy is emitted or absorbed can be
controlled by the chemical potentials $\mu_{\nu i}$ and mixing angles $\theta_{i}$.
Note further, that for $g_{0}=0$ these rates coincide with the corresponding rates
for the DQD:s decoupled from the cavity mode, as depicted in
Fig. \ref{exchange}.

Based on an analytical solution to the QME (See Appendix C), we
can further conclude that a
well defined steady-state solution for $\hat{\rho}$ exist for mixing angles $\theta_{1},\theta_{2}>\pi/4$. For smaller angles and certain bias configurations the photon
number can diverge. In Ref. \cite{Jin11}, where a single DQD-cavity system was
considered, it was shown that DQD mixing angles $\theta<\pi/4$ can lead to population inversion and hence a cavity lasing state. Since the
focus on this work is the few-photon regime we, if not otherwise explicitly
stated, focus on mixing angles $\theta_{1},\theta_{2}>\pi/4$ below.
\section{Non-local transport properties}
\label{nonLocal}
The main purpose of this work is to investigate non-local transport properties 
due to exchange of photons between tunneling electrons in different
DQD:s. This investigation is carried out in the regime $\hbar\omega\gg k_{B}T\gg \hbar
g_{0}$. The experiments reported in
Refs. \cite{Frey11,Delbecq11,Frey112} were performed under these conditions. Moreover,
for $k_{B}T\gg \hbar
g_{0}$ the occupations of the leads do not change significantly over the energy
scale $\hbar g_{0}$, i.e. $f_{\nu i}[\pm\hbar\omega/2+\mathcal{O}(\hbar
g_{0})]\approx f_{\nu
  i}(\pm\hbar\omega/2)$. Then there are effectively
only two energies in the leads at which the electrons can tunnel into and out
of the DQD:s. This simplifies the expressions for the tunneling rates into and out of the DQD:s (See
Sec. \ref{derQME}) and allows to reduce the QME to an ordinary master
equation (ME) (See Appendix C). We write this ME $d\mathbf{P}/dt=M\mathbf{P}$, where $\mathbf{P}$
is a vector with the probabilities of eigenstates of $\hat{H}_{S}$
and $M$ is the matrix with the transition rates between these eigenstates.

For the system to display non-local transport effects 
it is clear that two conditions must be fulfilled: First, transport electrons in different DQD:s
must exchange photons. Second, for at least one of the DQD:s the effective
tunneling rate into the empty DQD or out of the occupied DQD must be dependent
on the number of excitations in the system or the occupation of the other DQD. The most clear non-local effects will thus occur for a
state-dependence such that transport becomes blocked in one of the DQD:s if photons are not emitted by
the other. Here we take the DQD2 to be blocked when there are no
excitations in the system. This is accomplished by choosing a bias
configuration similar to the one depicted in Fig. \ref{exchange}, i.e. such that
$f_{L2}(-\hbar\omega/2)=1$, $f_{R2}(\hbar\omega/2)=0$, $f_{L2}(-\hbar\omega/2)=f_{L2}(\hbar\omega/2)=1$.

\subsection{Current}
\begin{figure}[ht]
\begin{center}
\includegraphics[trim=0cm 0cm 0cm 0cm, clip=true, width=0.40\textwidth,angle=0]{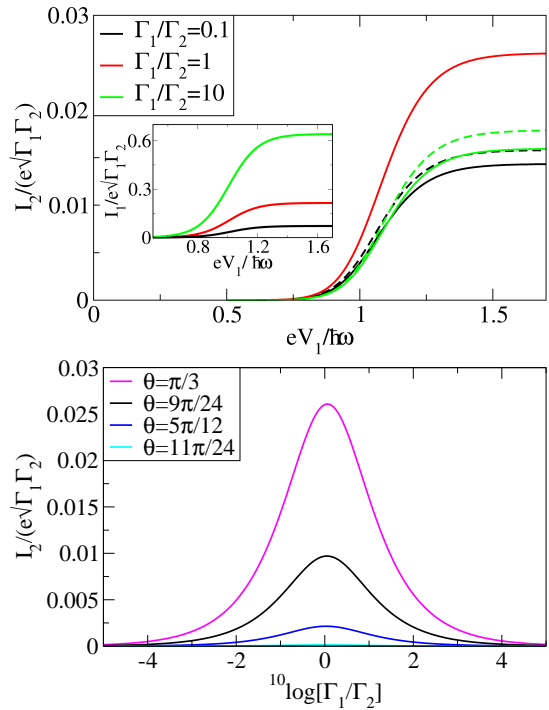}
\includegraphics[trim=0cm 0cm 0cm 0cm, clip=true, width=0.40\textwidth,angle=0]{resultcurr2mod.eps}
\caption{Upper panel: The non-locally induced current $I_{2}$ as a function of
  $eV_{1}/\hbar\omega$ for $\theta_{1}=\theta_{2}=\pi/3$,
  $k_{B}T=0.05\hbar\omega$ and different asymmetry factors
  $\Gamma_{1}/\Gamma_{2}$. The dashed black(green) curve shows the current for $\Gamma_{1}/\Gamma_{2}=1/10(10)$ obtained from the analytical expression for $\Gamma_{1}\ll\Gamma_{2}(\Gamma_{2}\ll\Gamma_{1})$ in Eq. \eqref{nonlocal}. Inset shows
  $I_{1}$ as a function of $eV_{1}/\hbar\omega$ for the same asymmetry factors and mixing
  angles. Lower panel: The current $I_{2}$ above the onset voltage against the
  asymmetry factor $\Gamma_{1}/\Gamma_{2}$ for
  different $\theta=\theta_{1}=\theta_{2}$.\label{resultcurr}} 
\end{center}
\end{figure}

We first consider the currents in the DQD:s as a function of the bias
voltage $V_{1}$ across DQD1. The current $I_{i}$ through DQDi is determined by
the populations of the eigenstates of the system and the effective tunneling rates
between the DQD and its right lead. The currents $I_{1}(V_{1})$ and
$I_{2}(V_{1})$ are plotted for symmetric bias, $\mu_{L1}=-\mu_{R1}=eV_{1}/2$, and $\theta_{1}=\theta_{2}=\pi/3$ in
Fig. \ref{resultcurr}. The key feature of both current-voltage characteristics
is a thermally broadened onset at $eV_{1}=\hbar\omega$. 
In DQD1 the onset occurs when the energies at which the electron can tunnel
into and out of the DQD enters the bias window. The electron tunneling through
DQD1 will further excite the system to states where tunneling out of DQD2
becomes possible. Hence the onset of the current $I_{2}$ occurs at the
same bias voltage $V_{1}$. In the limits
$\Gamma_{1}/\Gamma_{2}\ll 1$ and $\Gamma_{1}/\Gamma_{2}\gg 1$ the solution to
the ME, and hence the currents, can be obtained analytically (See Appendix
C). Focusing on the non-locally induced current $I_{2}(V_{1})$, for symmetric bias and $eV_{1}\gg k_{B}T$ we obtain
\begin{align}
&I_{2}=\frac{e\Gamma_{1}\gamma^{2}}{1+\gamma},\quad\Gamma_{1}/\Gamma_{2}\ll 1\\
&I_{2}=\frac{e\Gamma_{2}\gamma^2}{[1+2\cot^{2}(\theta_{2})](1-2\gamma+2\gamma^{2})+\gamma^{2}},\quad\Gamma_{1}/\Gamma_{2}\gg 1,\nonumber
\label{nonlocal}
\end{align}
where $\gamma=f_{L1}(\hbar\omega/2)\cos^{2}(\theta_{1})$. From these expressions it is clear that the magnitude of the induced
current can be made $\sim e\Gamma_{1}$ and $\sim e\Gamma_{2}$ in the limits
$\Gamma_{1}/\Gamma_{2}\ll 1$ and $\Gamma_{1}/\Gamma_{2}\gg 1$, respectively.
From the plot of the high-bias current against the asymmetry factor
$\Gamma_{1}/\Gamma_{2}$  for $\theta_{1}=\theta_{2}$ in Fig. \ref{resultcurr}
and from further investigation for $\theta_{1}\neq\theta_{2}$, we find 
that the non-local effect is maximal for $\Gamma_{1}\sim\Gamma_{2}$. We can
can thus conclude that the non-locally induced current,
qualitatively behaving as $I_{2}\sim
e\Gamma_{1}\Gamma_{2}/(\Gamma_{1}+\Gamma_{2})$, is considerable for a large range of the parameters $\Gamma_{1},\Gamma_{2},\theta_{1}$ and $\theta_{2}$.

To estimate the magnitude of the non-locally induced current we first recall that strong
coupling implies the limit
$\Gamma_{i}\ll g_{0}$ for the tunneling rates. In
recent single DQD-cavity experiments \cite{Frey112,Petersson12} fundamental frequencies $\omega/2\pi\sim 10$GHz and DQD-cavity
coupling strengths $g_{0}\sim 500$MHz were reported. This means that strong
coupling requires $\Gamma_{i}\lesssim 100$MHz giving non-locally induced currents 
of the order $I_{2}\sim 0.1$pA. Importantly, currents of this magnitude have
been measured in DQD-cavity systems \cite{Frey112}.

To further put the magnitude of the non-locally induced current in perspective we briefly
discuss non-local transport properties for the system with the DQD:s
replaced by single-level, or metallic dots. As follows from our
Ref. \cite{Bergenfeldt12}, the ME:s describing the evolution of these systems are explicitly dependent on the
parameter $Z_{0}/R_{Q}$, typically much smaller than unity for regular
transmission lines. Importantly, the ME:s show that the effective tunneling
rate into an empty dot or
out of an occupied dot is independent on the system state to first order in
$Z_{0}/R_{Q}$. It thus follows that the non-locally induced current will
be proportional to $(Z_{0}/R_{Q})^{2}$, to first non-vanishing order. In a
single-level dot- or metallic dot-cavity system the
non-locally induced current will thus, in contrast to the current in a DQD-cavity system, be significantly suppressed.

\subsection{Current correlations}
\begin{figure}[ht]
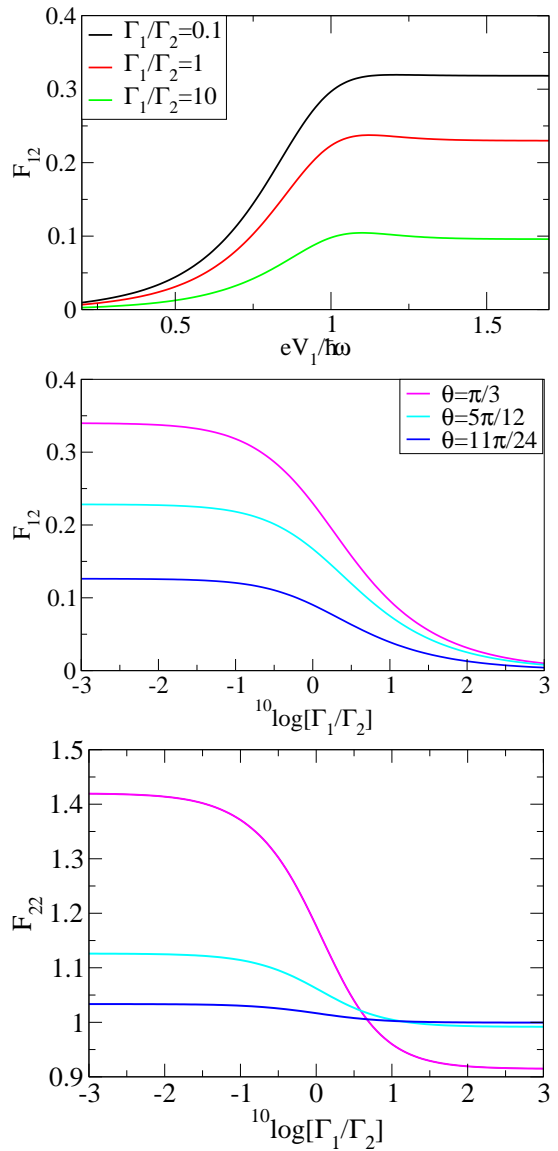

\begin{center}
\includegraphics[trim=0cm 0cm 0cm 0cm, clip=true, width=0.40\textwidth,angle=0]{Crossbiasmod.eps}
\includegraphics[trim=0cm 0cm 0cm 0cm, clip=true,
width=0.40\textwidth,angle=0]{results.eps}
\includegraphics[trim=0cm 0cm 0cm 0cm, clip=true,
width=0.40\textwidth,angle=0]{fano.eps}
\caption{Upper panel: Cross-correlation Fano factor $F_{12}=S_{12}/(e\sqrt{I_{1}I_{2}})$ a function of $eV_{1}$ for different asymmetry factors
  $\Gamma_{1}/\Gamma_{2}$ for mixing angles $\theta_{1}=\theta_{2}=\pi/3$ and
  temperature  $k_{B}T=0.05\hbar\omega$.  
Middle panel: Cross-correlation Fano factor above onset as a function of asymmetry parameter
$\Gamma_{1}/\Gamma_{2}$ for $\theta_{1}=\theta_{2}=\theta$ and $k_{B}T\ll\hbar\omega$.
Lower panel: Auto-correlation Fano factors $F_{22}=S_{22}/I_{2}$ above onset as a function of asymmetry parameter for
mixing angles $\theta_{1}=\theta_{2}=\theta$ and $k_{B}T\ll\hbar\omega$.\label{resultnoise}}
\end{center}
\end{figure}

Having established that photon exchange between transport electrons in the
spatially separated DQD:s can result in a non-local
current we, as the next natural step, investigate the mechanism behind this exchange. 
To this aim we consider the low-frequency correlations $S_{ij}$ between the
currents in DQDi and DQDj. Current correlations are known to provide
information about e.g. the effective charge, interactions and statistical
peorperties of the
charge carriers \cite{Blanterrev}. The correlations $S_{ij}$ can formally be obtained from a number-resolved version of the ME (See Appendix D). Focusing first on the cross-correlations $S_{12}$  we plot in Fig. \ref{resultnoise} 
the cross-correlation Fano-factor $F_{12}=S_{12}/(e\sqrt{I_{1}I_{2}})$ against the
bias voltage $V_{1}$ for different asymmetry factors. Similar to the
normalized currents
$I_{1}$ and $I_{2}$ the cross-correlations have an onset at
$eV_{1}=\hbar\omega$. However, in contrast to the current $I_{2}$ (but similar
to $I_{1}$) the cross-correlations have a strong dependence on the asymmetry factor $\Gamma_{1}/\Gamma_{2}$. This can be seen by considering the limits $\Gamma_{1}/\Gamma_{2}\ll
1$ and $\Gamma_{1}/\Gamma_{2}\gg 1$ where analytical expressions can be
obtained. Above the onset voltage we get
\begin{align}
&F_{12}=\frac{\cos(\theta_{1})[1+\cos^{4}(\theta_{1})]}{[1+\cos^{2}(\theta_{1})]^{2}},\quad \Gamma_{1}/\Gamma_{2}\ll 1\nonumber\\
&F_{12}=\mathcal{O}(\sqrt{\Gamma_{2}/\Gamma_{1}}),\quad \Gamma_{2}/\Gamma_{1}\ll 1. 
\end{align}
From these expressions we see that the currents in DQD1 and DQD2 are manifestly
positively correlated, $F_{12}>0$ ($\cos(\theta_{1})>0$), for $\Gamma_{1}/\Gamma_{2}\ll 1$.
The correlations are also strong, $F_{12}\sim 1$. In contrast, for
$\Gamma_{1}/\Gamma_{2}\gg 1$, the currents are essentially uncorrelated. The cross-over between the two
regimes is shown in Fig. \ref{resultnoise} for $\theta_{1}=\theta_{2}$.
The strong, positive correlations appearing for $\Gamma_{1}\lesssim\Gamma_{2}$ clearly
show that tunneling through DQD2 is triggered by tunneling through DQD1. 
The qualitatively different system behavior in the limits $\Gamma_{1}/\Gamma_{2}\ll 1$
and $\Gamma_{1}/\Gamma_{2}\gg 1$, respectively is also manifested in the auto-correlations $S_{22}$. In Fig. \ref{resultnoise} we see
that the auto-correlation Fano-factor $F_{22}=S_{22}/(eI_{2})$ above onset
goes from a sub-Poissonian value, $F_{22}<1$, to a super-Poissonian value, $F_{22}>1$, as the
asymmetry factor $\Gamma_{2}/\Gamma_{1}$ is decreased from infinity to zero.
This describes a transition from anti-bunching to bunching behavior of the
transport electrons \cite{Blanterrev}.

\begin{figure}[ht]
\begin{center}
\includegraphics[trim=0cm 0cm 0cm 0cm, clip=true, width=0.45\textwidth, angle=0]{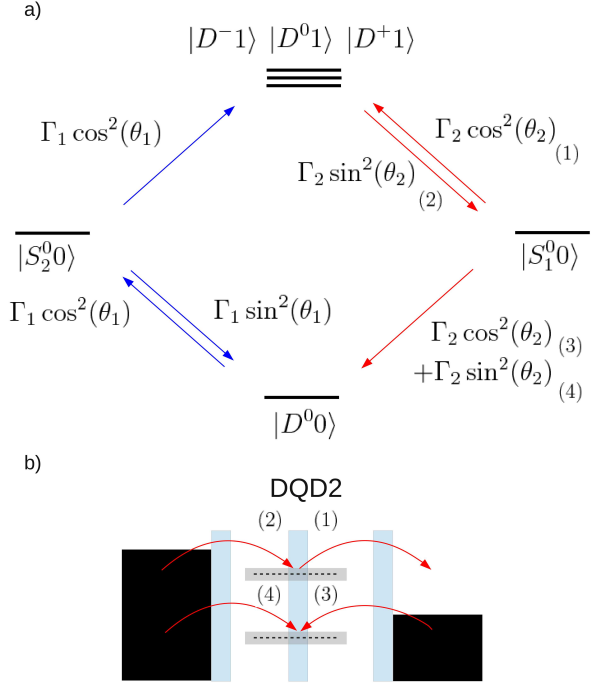}
\end{center}
\caption{(a) Schematic of processes in ME $d\mathbf{P}/dt=M\mathbf{P}$
  contributing to tranbsport quantities to first
  order in the asymmetry parameter $\Gamma_{1}/\Gamma_{2}$. The tunneling
  processes in DQD1 (slow processes) are marked by blue arrows and the
  tunneling processes in DQD2 (fast processes) are marked by red arrows.
  (b) Tunneling processes in DQD2. Process (1) and (2) describe
the tunneling into and out of the DQD as the system goes back-and forth
between $\ket{S_{1}^{0}0}$ and $\{\ket{D^{\alpha}1}\}$. Processes (3) and (4)
describe the tunneling processes where the system relaxed from
$\ket{S_{1}^{0}0}$ to $\ket{D^{0}0}$.
\label{scheme}}
\end{figure}

To connect these findings to the properties of the photon exchange we perform
a careful investigation of the ME in the limit $\Gamma_{1}\ll\Gamma_{2}$. The
processes contributing to  to transport quantities to leading order in the asymmetry parameter $\Gamma_{1}/\Gamma_{2}$ are depicted in Fig. \ref{scheme}. From this scheme it is apparent that the states $\ket{D^{0}0}$ and
$\ket{S_{2}^{0}0}$ will have occupations $\mathcal{O}(1)$, while the other states have occupation
$\mathcal{O}(\Gamma_{1}/\Gamma_{2})$. The system will thus spend most of its
time in the states $\ket{D^{0}0}$ and
$\ket{S_{2}^{0}0}$ and will
occasionally be excited out of this subspace by a tunneling event in DQD1, from
the state $\ket{S_{2}^{0}0}$ to any of the states
$\{\ket{D^{\alpha}1}\}$. The system can from here go back and forth 
between $\{\ket{D^{\alpha}1}\}$ and $\ket{S_{2}^{0}0}$ an arbitrary number of
times before relaxing to $\ket{D^{0}0}$. This will occur on a time-scale $\sim
1/\Gamma_{2}$. Each tunneling event in DQD1 which
excites the system to  $\{\ket{D^{\alpha}1}\}$ will hence be followed
by one or more tunneling events in DQD2 during a short time-window $\sim
1/\Gamma_{2}$. The electrons in DQD2 are thus transported in cascades induced by randomly occurring tunneling events in
DQD1 with separation $1/\Gamma_{1}\gg 1/\Gamma_{2}$. 
This mechanism, commonly referred to as
dynamical channel blockade \cite{Cottet04,Belzig05}, explains both the positive cross-correlations,
$S_{12}>0$, and the bunching of electrons in DQD2, $F_{22}>1$. 
Importantly, each cascade in DQD2 is initiated by the emission of a single
photon emitted by an electron tunneling through DQD1. Our investigation thus supports the physical picture where single photons are exchanged between
the DQD:s for $\Gamma_{1}\ll\Gamma_{2}$. 

\subsection{Full counting statistics}
To obtain a complete picture of the elementary processes of the charge transport
for $\Gamma_{1}\ll\Gamma_{2}$ we consider the full transport statistics. The
statistics is most clearly visualized via the cumulant
generating function (CGF) $\mathcal{F}$ which can be obtained analytically
above onset (See Appendix D) and is given by
\begin{eqnarray}
&\mathcal{F}(\chi_{1},\chi_{2})=-\frac{\Gamma_{1}}{2}\Bigg(1+\cos^{2}(\theta_{1})\nonumber\\
-&\sqrt{\sin^{4}(\theta_{1})+4\cos^{2}(\theta_{1})e^{i\chi_{1}}(\sin^{2}(\theta_{1})+\cos^{2}(\theta_{1})y)}\Bigg)
\nonumber\\
&y=\left(\cos^{2}(\theta_{2})+\sin^{2}(\theta_{2})e^{i\chi_{2}}\right)\sum_{n=0}\frac{z^{n}}{z_{0}}e^{in\chi_{2}}.
\label{CGF}
\end{eqnarray}
Here $z=\cos^{2}(\theta_{2})/[1+\cos^{2}(\theta_{2})]$,
$z_{0}=1+\cos^{2}(\theta_{2})$ and $\chi_{i}$ is the counting field for charge
transfer in DQDi.
To interpret the CGF we first consider the case when the charge
transfer through DQD2 is not monitored, i.e. $\chi_2=0$. Then $y=1$
and the CGF reduces to the well known result \cite{Bagrets03} for a single
level dot with tunneling rates $\Gamma_1$ and
$\Gamma_1\cos^2(\theta_1)$ into and out of the dot, respectively. For
$\chi_2 \neq 0$, it is clear from the term
$\sin^2(\theta_1)+\cos^2(\theta_1)y$ that tunneling events into DQD1 are of two
kinds. First, for events corresponding to the transition
$\ket{S_{2}^{0}}\rightarrow\ket{D^{0}0}$, occurring with a rate
$\propto \sin^2(\theta_1)$, there is no tunneling in DQD2. 
The second type of events, corresponding to transitions
$|S_2^00\rangle \rightarrow \{|D^{\alpha} 1\rangle\}$ and occurring
with a rate $\propto \cos^2(\theta_1)$, trigger tunneling in
DQD2. This tunneling takes place as the system goes back to
$\ket{D^00}$ and is described by the function $y=y(\chi_2)$.

The different terms in $y$ are given by the probabilities for all
possible processes taking the system from $\{|D^{\alpha} 1\rangle\}$
to $\ket{D^{0}0}$, weighted with counting field factors describing the
respective charge transfer through DQD2. Common for all processes is
that they start with the transition $\{\ket{D^{\alpha} 1}\}
\rightarrow \ket{S_1^00}$ and end with $\ket{S_1^00} \rightarrow
\ket{D^00}$. These two transitions give rise to the prefactor
$\cos^2(\theta_2)+\sin^2(\theta_2)e^{i\chi_2}=e^{i\chi_2}[\cos^2(\theta_2)e^{-i\chi_2}+\sin^2(\theta_2)]$,
with the $e^{i\chi_2}$ from the starting transition and
$\cos^2(\theta_2)e^{-i\chi_2}+\sin^2(\theta_2)$ from the ending
transition. As shown in Fig. 6, the ending transition can occur via
tunneling from the left lead of DQD2 (probability $\sin^2(\theta_1)$, no counting
field factor) or back from the right lead DQD2 (probability $\cos^2(\theta_1)$,
counting field factor $e^{-i\chi_2}$). The sum term in $y$, running
from zero to infinity, describes all possible back-and-forth
transition $\ket{S_1^00} \rightarrow \{\ket{D^{\alpha} 1}\}
\rightarrow \ket{S_1^00}$ the system can perform, between the
starting and ending transitions (see Fig 6). The $n$:th term in the sum
thus corresponds to $n$ transitions and hence $n$ electrons being transferred
across DQD2. This interpretation comes naturally when noting that 
$z^{n}/z_{0}$ is the probability to return to the state $\ket{D^{\alpha} 1}$
$n$ times. The structure of the CGF, and in particular $y$, clearly shows that
that electrons in DQD2 are transported in cascades and that these cascades are
triggered by single photons emitted by electrons tunneling through DQD1. 

\section{Spectral fine structure and entanglement}
\label{TavisCummingsfine}
It is interesting to investigate what qualitatively new physical effects come into
play in the regime where the thermal broadening in the leads is
much smaller than the DQD-cavity coupling strength, i.e. $k_BT \ll \hbar g_{0}$. In this regime the ME-description used in the previous section is no
longer valid and we need to consider the full QME of Eq. \eqref{QME}.

\subsection{Transport properties}
 
We first demonstrate that the structure on the scale $\sim\hbar g_{0}$ in
the spectrum of the generalized TC Hamiltonian appear in the transport
properties of the system for $k_BT \lesssim \hbar g_{0}$. This fine
structure, a manifestation of coherent electron-photon interaction, appear already in the average current and we
therefore focus on this quantity. Moreover, we consider the
simplest possible parameter regime by taking DQD:s with identical mixing
angles $\theta_{1}= \theta_{2}=\theta$, lead-DQD tunneling rates
$\Gamma_{1}=\Gamma_{2}=\Gamma$ and with left (right) leads having the same
chemical potential, i.e. $\mu_{\nu 1}=\mu_{\nu 2}=\mu_{\nu}$. This will give
the same current $I=I_{1}=I_{2}$ in DQD1 and DQD2. This current $I$ is readily obtained from a numerical solution of
Eq. \eqref{QME} (See Appendix D for details). In Fig. \ref{fintemp} we plot
$I(V)$ for symmetric bias voltage $\mu_{L}=-\mu_{R}=eV/2$ and $\theta=\pi/3$
for temperatures ranging from $k_BT \sim \hbar g_{0}$ down to
$k_BT \ll \hbar g_{0}$. We clearly see how the
single step onset at $eV=\hbar\omega$ is split up into several smaller steps,
spaced $\sim\hbar g_{0}$ as the temperature is decreased. These steps can
directly be attributed to the structure of the spectrum of the generalized TC
Hamiltonian. They are a consequence of eigenstates with energy splittings
$\sim\hbar g_{0}$ becoming populated at different bias voltages. 
It is here interesting to note that signatures of the JC spectrum was found in the
frequency-dependent current auto-correlations in the transport through a
system with only one DQD coupled to the cavity mode \cite{Lambert09}.   
\begin{figure}[ht]
\begin{center}
\includegraphics[trim=0cm 0cm 0cm 0cm, clip=true, width=0.40\textwidth, angle=0]{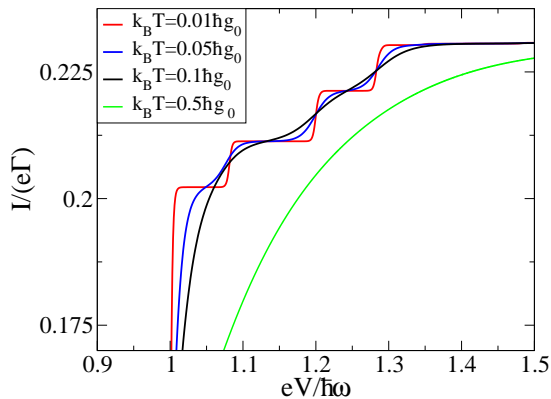}
\caption{The current $I=I_{1}=I_{2}$ through the DQD:s as a function of
  voltage above onset for $\theta=\pi/3$, $g_{0}=0.1\omega$ and different temperatures.\label{fintemp}}
\end{center}
\end{figure}

\subsection{Transport-induced entanglement}

\begin{figure}[ht]
\begin{center}
\includegraphics[trim=0cm 0cm 0cm 0cm, clip=true, width=0.40\textwidth,
angle=0]{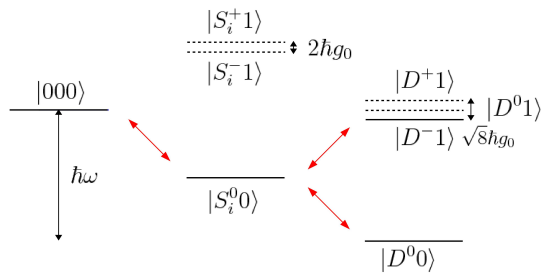}
\caption{Scheme of the lowest energy levels in the generalized TC Hamiltonian.
The red arrows show the active transitions for $k_{B}T\ll\hbar g_{0}$ and for a bias configuration such that the chemical potentials satisfy the conditions in Eq \eqref{biasconf}. \label{diagcon}}
\end{center}
\end{figure}

A natural question to ask when considering two coupled, spatially
separated, DQD:s is to what extent their orbital degrees of freedom become entangled by
the exchange of cavity photons. The object of interest, describing the
properties of the electronic state with one electron in each DQD, is
the reduced two-particle density matrix $\hat{\rho}_{r}$. The reduced
density matrix, of dimension $4\times 4$, is formally obtained by
first projecting the total system density matrix $\hat{\rho}$ onto the
TC subspace and then tracing out
the photonic degrees of freedom. As follows from the structure of
Eq. \eqref{QME} and the TC-eigenstates (Appendix B), the reduced density
matrix can be written as a sum of the four diagonal components in the singlet-triplet basis,
\begin{equation}
\hat{\rho}_{r}=\rho_{g}|gg\rangle \langle gg|+\rho_{e}|ee\rangle \langle
ee|+\rho_S|\mathcal{S}\rangle
\langle \mathcal{S}|+\rho_T|\mathcal{T}\rangle \langle \mathcal{T}|
\label{rhor}
\end{equation}
where $\ket{\mathcal{S}(\mathcal{T})}=(\ket{eg} -(+)\ket{ge})/\sqrt{2}$. The entanglement
of $\hat{\rho}_{r}$ is conveniently quantified via the concurrence
\cite{Wooters98}, ranging from $1$ for a maximally entangled state to
$0$ for a non-entangled, separable state. For a density matrix on the
form in Eq. \eqref{rhor} the concurrence $C(\hat{\rho}_{r})$ takes on
the simple form
\begin{equation}
C=\mbox{max}\{|\rho_{S}-\rho_{T}|-2\sqrt{\rho_{e}\rho_{g}},0\}
\label{conc}
\end{equation}
To determine if entanglement can be induced by photon exchange we
first consider the scheme in Fig. \ref{diagcon}, displaying the
lowest energy states with the TC-subspace well resolved. By noting that
the lowest excited TC-states $|D^{\alpha}1\rangle$ are written 
$|D^01\rangle=|\mathcal{S}0\rangle$ and
$|D^{\pm}1\rangle=(1/\sqrt{2})(|gg1\rangle \pm |\mathcal{T}0\rangle)$ it is
clear that a selective population of any of the $|D^{\alpha}1\rangle$
states would give an electronic state with a large singlet ($\mathcal{S}$) or
entangled triplet ($\mathcal{T}$) component. To demonstrate such a selective
population we choose bias voltages $V_{i}$ and
dot-level positions such that the chemical potentials $\mu_{i
  L}$ and $\mu_{i R}$ obey the
relations
\begin{align}
&\frac{\hbar\omega}{2}-\sqrt{2}\hbar g_{0}<\mu_{Li}<\frac{\hbar\omega}{2}-\hbar g_{0},\nonumber\\
&-\frac{\hbar\omega}{2}>\mu_{Ri}>-\frac{\hbar\omega}{2}-(\sqrt{2}-1)\hbar g_{0}.
\label{biasconf}
\end{align}
For $k_{B}T\ll\hbar g_{0}$ then only
$\ket{D^{-}1}$ of the excited TC-states becomes populated.

For the chosen parameters, as seen in Fig. \ref{diagcon}, only five states of the generalized TC-model
contributes to transport. For this case the QME can be solved
exactly. Importantly, the steady-state solution gives a reduced two-particle density
matrix $\hat{\rho}_{r}$ with only $\rho_{g}$ and $\rho_{T}$
nonzero. As is clear from Eq. \eqref{conc} the resulting concurrence is
finite. For a symmetric
parameter setting, i.e. for $\theta_{1}=\theta_{2}=\theta$,
$\Gamma_{1}=\Gamma_{2}=\Gamma$ and $\mu_{\nu 1}=\mu_{\nu 2}=\mu_{\nu}$ this concurrence is
given by 
\begin{equation}
C=\frac{\cos^{4}(\theta)}{2[\cos^{4}(\theta)+\sin^{4}(\theta)]},
\label{conc2}
\end{equation}
showing that the concurrence can reach up to $C=1/2$ for $\theta \ll 1$. We stress
that for the chosen parameters there is no bound on $\theta$ in order
to have a well defined solution to the QME.

\begin{figure}[h]
\begin{center}
\includegraphics[trim=0cm 0cm 0cm 0cm, clip=true, width=0.40\textwidth,
angle=0]{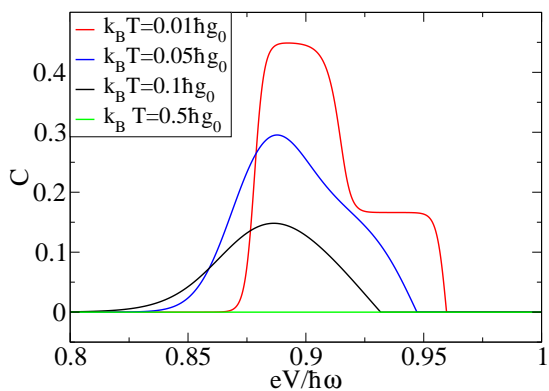}
\caption{Concurrence 
$C$ as a function of bias voltage for different temperatures
  for $g_{0}=0.1\omega$, $\mu_{L}+\mu_{R}=-0.12\hbar\omega$
and $\theta_{1}=\theta_{2}=\pi/6$.
\label{concurrence}}
\end{center}
\end{figure}

Having confirmed the existence of large entanglement $C\lesssim 1/2$
we consider the effect of finite temperature and modified bias voltage
$V=V_{1}=V_{2}$. Solving numerically the QME, the resulting concurrence $C(V)$ is
plotted in Fig. \ref{concurrence} for different temperatures. For low
temperatures $k_BT\ll \hbar g_0$ the entanglement has an onset when
$\ket{D^{-}1}$ is populated, with the concurrence given by
Eq. (\ref{conc2}). Increasing the voltage further $\ket{D^{0}1}$ is
populated as well, decreasing the concurrence due to the finite
probability for both entangled triplet, $\rho_{T}$ and singlet,
$\rho_{S}$, electronic states, clear from Eq. \eqref{conc}. For even
larger bias all TC-states $|D^{\alpha}1\rangle$ have finite population
and the entanglement disappears. From Fig. \ref{concurrence} it is
also clear that increasing the temperature smears the $C(V)$ curve and
successively suppresses the entanglement, reaching a separable state at
$k_BT \approx \hbar g_0/2$.

We can thus conclude that both in the high-bias and high-temperature
regimes, where the system can be described by a ME, the entanglement
is zero. To clarify the generality of this observation we investigated
the concurrence in all regimes where we could solve the ME
analytically (see Appendix C). In these regimes we could formally
prove the absence of entanglement. Moreover we considered the
concurrence obtained numerically for a broad range of other system
parameters in the ME-regime but did not find any entanglement.  We
thus conclude that it is highly probable that electrons in the two
DQD:s can only be entangled for temperatures $k_BT \ll \hbar g_{0}$, in
biasing regimes where TC-states with the same number of excitations
are selectively populated.

We stress that our investigation mainly aims at demonstrating the
existence of entanglement. We do not analyze how or even if it can be
detected by transport measurements. Moreover, we have not made a full
investigation over the entire parameter space to identify the regime
with the largest entanglement.

\section{Dephasing and relaxation effects}
\label{Exp}
So far we have neglected dephasing and relaxation effects
in the DQD:s as well as loss of cavity
photons. From the recent single DQD experiments
\cite{Frey112,Petersson12} it is clear that the dephasing rate
$\Gamma_D$ is much larger than the rates $\Gamma_R$ and $\kappa$ for
relaxation and cavity loss, respectively. We thus focus on the effect
of dephasing on the results presented above.

Dephasing can qualitatively be accounted for by adding a term \cite{Nielsen10}
\begin{equation}
\mathcal{L}_{D}[\hat{\rho}]= \frac{\Gamma_D}{2}\sum_{i=1,2}\left[2 \hat L_i^{\dagger} \hat \rho
   \hat L_i-\hat L_i^{\dagger}\hat L_i\hat \rho-\hat \rho \hat 
L_i^{\dagger}\hat L_i\right]
\label{deph}
\end{equation}
to the Liouvillian in Eq. (\ref{QME}). Here $\hat L_i=\hat d_{ei}^{\dagger} \hat
d_{ei}-\hat d_{gi}^{\dagger} \hat d_{gi}$ and the dephasing is taken
independent, with the same rate $\Gamma_D$, for the two DQD:s. An investigation of dephasing in all parameter regimes is beyond
the scope of the present article. However, we stress that for strong
dephasing, $\Gamma_D \gg g_0$, coherent superpositions between excited
and ground states in the DQD:s are suppressed. The steady state solution of 
Eqs. (\ref{QME}) and
  (\ref{deph}) is diagonal in the basis of the DQD-cavity product states 
$\ket{\xi_{1}\xi_{2} p}$,
with $\xi _{i}=0,g,e$ and $p$ the number of photons. As a consequence, 
electrons
and photons are decoupled and the non-local transport effects as well as the DQD
entanglement appearing in the regime $\hbar g_{0}\gg k_{B}T$ are suppressed.

Importantly, in both single DQD-experiments \cite{Frey112,Petersson12}
the dephasing is found to be strong, with $\Gamma_D\sim 1$ GHz, 
substantially larger
than the coupling strength $g_0/2\pi\sim 100$ MHz. It is thus necessary to
consider ways to increase $g_0$ and/or
suppress $\Gamma_D$, in order to approach the
strong coupling limit $g_0 \gg \Gamma_D$ where the non-local effects
discussed above are fully developed. First and foremost, the
coupling $g_0$ can be increased substantially by increasing the 
fundamental
frequency ($g_0\propto \omega$), simply by making a shorter cavity. 
Importantly,
since we consider an isolated cavity, $\omega$ is not limited by
requirements of an external microwave circuitry. The limit is
instead set by the energy gap of the superconducting cavity material, of 
the order of
hundreds of GHz for large gap superconductors as e.g. Nb ($\omega/2\pi \approx 
10$
GHz in \cite{Frey112,Petersson12}). Second, unconventional
transmission line cavities, with a central
conductor consisting of e.g. Josephson junctions or SQUIDS,
\cite{Castellanos07,Castellanos08,Masluk12} can have characteristic impedances
$Z_0 \sim 1$k$\Omega$. This gives coupling strengths $g_0 \sim 0.1
\omega$, one order of magnitude larger than for conventional
transmission lines. Third, since Refs. \cite{Frey112,Petersson12}
are the first experiments on DQD:s in cavities, there is probably room for
for optimizing the circuit design, further suppressing the dephasing. 
Taken
together, the strong coupling limit of our proposal is
arguably within reach experimentally. Moreover, the relaxation rate in 
\cite{Frey112}
was estimated to $\Gamma_R \sim 100$ MHz, one order of magnitude 
smaller
than the dephasing rate. Hence, in the strong coupling limit $g_0 \gg 
\Gamma_D$,
relaxation is expected to be negligible. In addition, the cavity loss rate 
$\kappa$
in \cite{Frey112,Petersson12} was already much smaller than $g_0$,
suggesting that cavity loss can safely be neglected.

\section{Conclusion}
In conclusion we have theoretically investigated the
non-local transport properties of a DQD-cavity system.
We have found that the photons emitted by electrons
tunneling in one DQD can assist transport of electrons
through the other DQD, giving a strong non-locally induced
current and large cross-correlations
between currents in the two DQD:s. Moreover, in the
low temperature regime, $k_{B}T \ll\hbar g_{0}$, we have demonstrated
that signatures of the TC-spectrum will
appear in the I-V characterstics and that the orbital degrees
of freedom of electrons in the two DQD:s can become entangled.
Importantly, our work provide a theoretical framework for investigations of
non-local electronic transport properties in cavity -coupled nanoscale
conductors. The analysis can readily be modified to study transport through
other nanoscopic two-level systems  coupled to cavities, e.g. superconducting
single electron transistors \cite{Astafaiev07,Marthaler08} and spin qubits
\cite{Trif08,Cottet10,Jin12,Petersson12}

\section*{Acknowledgments}
We thank Seigo Tarucha, Per Delsing, Daniel Karlsson, Olov Karlstr\"{o}m and Cecilia Jarlskog for fruitful
discussions and input. We would further like to thank Christian Flindt, Bj\"{o}rn
Sothmann, Tineke van der Berg and Martin Leijsne for constructive comments on
an earlier version of the manuscript. We also acknowledge
discussion with Deborah Contreras-Pulido on a manuscript in preparation on a
closely related problem. The work was supported by the Swedish VR. 

\section*{Appendix A}

We here derive the Hamiltonian for the DQD-transmission line cavity system.
The first step is to describe the DQD:s within the standard \cite{vanderViel02}
constant-interaction model. Then only the excess dot charges will interact
capacitively with the cavity. This means that the total Hamiltonian of the system, $\hat{H}_{S}$,
becomes the sum of the Hamiltonian for the orbital states of the DQD:s,
$\hat{H}_{O}$, and the Hamiltonian for the cavity, the dot
charges and their interactions, $\hat{H}_{C}$. 
For the orbital part $\hat{H}_{O}$ we consider DQD:s formed by two tunnel
coupled quantum dots with a single active spin-degenerate level in each
dot. The orbital part of the Hamiltonian, in the localized basis of the DQD:s, then has the form
\begin{align}
\hat{H}_{O}=\sum_{i=1,2}\frac{\Delta_{i}}{2}(\dup_{Li}\ddown_{Li}-\dup_{Ri}\ddown_{Ri})+t_{LRi}(\dup_{Li}\ddown_{Ri}+\dup_{Ri}\ddown_{Li})
\label{orbital}
\end{align}
with $\Delta_{i}$ being the energy difference between the bare energies of the orbitals in the left and right dot
of DQDi. We recall that $t_{LRi}$ denotes the interdot tunneling amplitude of
DQDi and note that the creations operators $\dup_{Li},\dup_{Ri}$
are related to the eigenbasis creation operators creation operators according to  
\begin{align}
&\dup_{Li}=-\sin(\theta_{i})\dup_{gi}+\cos(\theta_{i})\dup_{ei},\nonumber\\
&\dup_{Ri}=\cos(\theta_{i})\dup_{gi}+\sin(\theta_{i})\dup_{ei}.
\end{align}

\begin{figure}[h]
\begin{center}
\includegraphics[trim=0cm 0cm 0cm 0cm, clip=true, width=0.45\textwidth, angle=0]{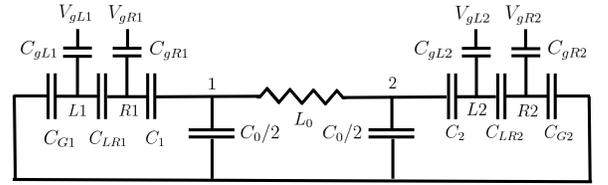}
\end{center}
\caption{Diagram of the circuit describing the transmission line cavity, the
  dot charges and their interactions. In DQD1 (DQD2) the right (left) dot is
  coupled capacitively to the central conductor of the transmission line,
  modelled by a LC-circuit and to two gate electrodes. The nodes $\nu i$ correspond to the dots while
  node 1 and 2 correspond to the endpoints of the transmission line.\label{circ}}
\end{figure}

The Hamiltonian $\hat{H}_{C}$ is derived within the framework of circuit QED. Following the procedure of
Refs. \cite{Devoret,Falci,Bergenfeldt12} we start from the classical Lagrangian
of a circuit representation of the system, including the capacitances of the
dots.  The transmission line is modelled by a single $LC$-circuit. This will
describe the physics of one finite-frequency (the fundamental mode)
\cite{devoret07} and the zero-frequency mode. 
The circuit diagram is shown in Fig. \ref{circ}, where
$C_{Gi}$, $C_{g\nu i}$, $C_{LRi}$ and $C_{i}$, $V_{g\nu i}$ denotes the
capcitances and gate voltages of DQDi and
$L_{0}$, $C_{0}$ are the total inductance and total capacitance to ground of the central
conductor. 

The Lagrangian of the circuit is given by
\begin{align} 
&\mathscr{L}=\sum_{i=1,2}\Bigg(\frac{C_{i}(\dot{\phi}_{i}-\delta_{i1}\dot{\phi}_{Ri}-\delta_{i2}\dot{\phi}_{Li})^{2}}{2}+\frac{C_{0}\dot{\phi}_{i}^{2}}{4}\nonumber\\
&+\frac{C_{LRi}(\dot{\phi}_{Li}-\dot{\phi}_{Ri})^{2}}{2}+\frac{\sum_{\nu}C_{g\nu i}(\dot{\phi}_{\nu
    i}-V_{g\nu i})^{2}}{2}
\Bigg)\nonumber\\
&+\frac{C_{G1}\dot{\phi}_{L1}^{2}+C_{G2}\dot{\phi}_{R2}^{2}}{2}-\frac{(\phi_{1}-\phi_{2})^{2}}{2L_{0}},
\label{circuitHam}
\end{align}
where $\phi_{\nu i}$ and $\phi_{i}$ denote the phases of node $\nu i$ and
$i$, respectively (see Fig. \ref{circ}). The zero- and finite-frequency normal modes, describing the
electrostatics and -dynamics of the circuit, respectively, are obtained from
the Euler-Lagrange equations. By rewriting the Lagrangian in terms of these
normal modes, performing a Legendre transformation and a canonical quantization of the 
fundamental mode a quantum Hamiltonian, $\hat{H}_{C}$, is
obtained. By further writing the excess charges of the dots as
$e(\dup_{\nu i}\ddown_ {\nu i}-n_{g\nu i})$, where $n_{g\nu i}$
denotes the gate-induced charges, we get
$\hat{H}_{C}=\hat{H}_{I}+\hat{H}_{DI0}+\hat{H}_{DI1}$, with 
\begin{align}
&\hat{H}_{I}=\hbar\omega\aup\adown\nonumber\\
&+\sum_{\nu i}\left[\frac{e^{2}(\dup_{\nu i}\ddown_{\nu i}-n_{g\nu i})^{2}}{2\tilde{C}_{\nu i}}+\lambda_{\nu i}(\aup+\adown)(\dup_{\nu i}\ddown_{\nu i}-n_{g\nu i})\right]\nonumber\\
&\hat{H}_{DI0}=\sum_{\nu\mu}U_{\nu 1\mu 2}(\dup_{\nu
    1}\ddown_{\nu 1}-n_{g\nu 1})(\dup_{\mu2}\ddown_{\mu2}-n_{g\mu2})\nonumber\\
&\hat{H}_{DI1}=\sum_{\nu\mu}\frac{\lambda_{\nu 1}\lambda_{\mu 2}}{2\hbar\omega}(\dup_{\nu
    1}\ddown_{\nu 1}-n_{g\nu1})(\dup_{\mu 2}\ddown_{\mu 2}-n_{g\mu2}),
\label{Hc}
\end{align}
In the total Hamiltonian $\hat{H}_{S}=\hat{H}_{C}+\hat{H}_{O}$, the part
$\hat{H}_{O}+\hat{H}_{I}$ has the standard form for a few-level
dot system linearly coupled to a bosonic mode (see e.g. Ref.\cite{Mitra04}),
with $\tilde{C}_{\nu  i}$ and $\lambda_{\nu i}$ denoting the effective self-capacitances
of the dots and the coupling strengths between the photons of the fundamental mode
and the dot charges, respectively.
The part $\hat{H}_{DI0}$ contains cross terms, $\sim\dup_{\nu 1}\ddown_{\nu
  1}\dup_{\mu 2}\ddown_{\mu 2}$, describing direct non-local coupling between the dot
charges. The coupling strengths, $U_{\nu 1\mu 2}$, depend parametrically only
on the capacitances of the circuit in Fig. \ref{circ}, i.e.  not on the inductance
$L_{0}$, and  would thus remain
unchanged if this inductance was short-circuited. This means that
$\hat{H}_{DI0}$ describes purely electrostatic, or capacitive, coupling. 
The part $\hat{H}_{DI1}$, just as $\hat{H}_{DI0}$, describes direct non-local coupling
between the dot charges. However, in contrast to $\hat{H}_{DI0}$ the coupling
strengths in $\hat{H}_{DI1}$, $\lambda_{\nu 1}\lambda_{\mu 2}/\hbar\omega$,
depend parametrically on the inductance $L_{0}$. This part is therefore electrodynamic.

As the next step we motivate the approximations leading from
Eqs. \eqref{orbital} and \eqref{Hc} to Eq. \eqref{Tavis}, under the conditions
described in the main text. To do this we use the relations
$\lambda_{\nu i}\propto\sqrt{Z_{0}/R_{Q}}\hbar\omega$ $(Z_{0}=\sqrt{L_{0}/C_{0}})$, $\omega\sim 1/\sqrt{L_{0}C_{0}}$
and $U_{\nu 1\mu 2}\propto e^{2}/C_{0}$ for the parameters in
$\hat{H}_{C}$. First, noting that $Z_{0}\ll R_{Q}$ and hence $\lambda_{\nu
  i}/\hbar\omega\ll 1$ justifies a rotating-wave 
approximation, which amounts to neglecting all terms of $\mathcal{O}(\lambda_{\nu
  i}/\omega)$ in $\hat{H}_{S}$, e.g. the counter-rotating terms.
Second, for the DQD:s resonant with the cavity mode the direct capacitive
interaction will scale as $U_{\nu 1\mu 2}/\lambda_{\nu (\mu) i}\sim
\sqrt{Z_{0}/R_{Q}}$ and can thus be neglected. 
Moreover, had we considered a full description for the transmission-line
DQD circuit, including all of the cavity modes, the higher frequency modes would be
off-resonant with a detuning $\Delta E\gtrsim\hbar\omega$. The corrections due
to this off-resonant interaction would then scale as $\lambda_{\nu
  i}/\hbar\omega$ and therefore be negligible.   
It should be noted that for the resonance condition to hold the tunneling
amplitudes and detunings between the left and right dot-orbitals the DQD:s 
must be chosen such that  
$|\Delta_{i}|=2\sqrt{(\hbar\omega)^{2}-t_{LRi}^{2}}$. 
We further point out that the DQDi-cavity coupling strengths is
$g_{i}=\sin(2\theta_{i})(\lambda_{Ri}-\lambda_{Li})/2$. Thus,
for the case of identical coupling strengths  $g_{1}=g_{2}=g_{0}$, considered
in the main text, the mixing angles $\theta_{1}$ and $\theta_{2}$ cannot be tuned
independently for fixed $\lambda_{Li}$ and $\lambda_{Ri}$.  

\section*{Appendix B}
We here give the explicit form for the eigenstates with a finite number of
excitations in the TC-subspace and all eigenenergies of $\hat{H}_{S}$. The former are given by
\begin{align}
&\ket{D^{0}1}=\ket{\mathcal{S}0},\quad\ket{D^{\pm}1}=\frac{\ket{gg1}\pm\ket{\mathcal{T}0}}{\sqrt{2}}
\end{align}
and for $n\geq 2$
\begin{align}
&\ket{D^{+-}n}=\frac{\sqrt{n-1}\ket{ggn}-\sqrt{n}\ket{een-2}}{\sqrt{2n-1}},\quad\nonumber\\
&\ket{D^{\pm\pm}n}=\frac{\sqrt{n}\ket{ggn}+\sqrt{n-1}\ket{een-2}}{\sqrt{2(2n-1)}}\pm\frac{\ket{\mathcal{T}n-1}}{\sqrt{2}},\nonumber\\
&\ket{D^{-+}n}=\ket{\mathcal{S}n-1},
\end{align}
with $\ket{\mathcal{S}(\mathcal{T})n}=(\ket{egn}-(+)\ket{gen})/\sqrt{2}$. The eigenenergies are
\begin{align}
&\epsilon_{00n}=\hbar\omega
(n+1)
\end{align} 
for the HO subspace,
\begin{align}
&\epsilon_{S^{0}_{i}0}=\hbar\omega/2,\hspace{2mm}\epsilon_{S^{\pm}_{i}n}=\hbar\omega(n+1/2)\pm\sqrt{n}\hbar g_{0},\hspace{2mm}n\geq 1
\end{align}
for the JC subspaces, and 
\begin{align}
&\epsilon_{D^{0}0}=0,\quad\epsilon_{D^{0}1}=\hbar\omega/2,\quad\epsilon_{D^{\pm}1}=\hbar\omega\pm\sqrt{2}\hbar g_{0},\nonumber\\
&\epsilon_{D^{\pm\pm}n}=n\hbar\omega\pm\sqrt{2(2n-1)}\hbar g_{0},\hspace{2mm}\epsilon_{D^{\pm\mp}n}=n\hbar\omega,\hspace{2mm}n\geq 2
\end{align}
for the TC subspace.

\section*{Appendix C}

In this Appendix we explain how the QME in Eq. \eqref{QME} can be reduced to a ME in the limit $\hbar g_{0}\ll k_{B}T$, give the explicit form of the
ME and solve it in three limiting cases. We start by pointing out that in the
secular regime $\Gamma_i \ll g_0$, considered here, only coherences between
degenerate states, i.e. $|+-n\rangle$ and $|-+n\rangle$ need to be taken into account in the QME.  Moreover, only the diagonal elements $\langle
S_i^{\alpha}n|\hat \rho |S_i^{\alpha}n\rangle$ couple to the
coherences. As pointed out in the text, for $\hbar g_0 \ll k_BT$, the
QME becomes independent on $g_0$. This introduces additional
symmetries in the QME, with two important consequences: $(1)$ The coherences $\langle +-n|\hat \rho |-+n\rangle$ and $\langle -+n|\hat
\rho |+-n\rangle$ couple with opposite signs to $\langle S_i^+n|\hat
\rho |S_i^+n\rangle$ and $\langle S_i^-n|\hat \rho |S_i^-n\rangle$.
$(2)$ For several pairs of diagonal elements of $\hat \rho$, only the
sums of the elements couple to the other diagonal elements. In
particular, this holds for the sum $\langle S_i^+n|\hat \rho
|S_i^+n\rangle+\langle S_i^-n|\hat \rho |S_i^-n\rangle$, to which the
coherences, according to (1), do not contribute. As a result of $(1)$ and $(2)$, the coherences decouple from the diagonal
elements of the QME, allowing us to reduce it to a standard ME. 

To write the explicit form of the ME it is 
convenient to first introduce a shorthand notation for the diagonal elements of $\hat \rho$, i.e. the
probabilities for the eigenstates of $\hat H_S$. The probabilities, or the
sums of probabilities, for states the HO, JC and TC subspaces are denoted by
\begin{align}
&P^{n}_{00}=\braket{00n|\hat{\rho}|00n},\nonumber\\ 
&P_{S_{i}}^{n}=\delta_{n0}\braket{S_{i}^{0}0|\hat{\rho}|S_{i}^{0}0}+(1-\delta_{n0})\sum_{\alpha=\pm}\braket{S_{i}^{\alpha}n|\hat{\rho}|S_{i}^{\alpha}n}
\end{align}
and
\begin{align}
&P_{D}^{n}=\delta_{n0}\braket{D^{0}0|\hat{\rho}|D^{0}0}+\delta_{n1}\sum_{\alpha=\pm}\braket{D^{\alpha}1|\hat{\rho}|D^{\alpha}1}\nonumber\\
&+(1-\delta_{n0}-\delta_{n1})\sum_{\alpha=\pm}\braket{D^{\alpha\alpha}n|\hat{\rho}|D^{\alpha\alpha}n},\nonumber\\
&P_{D^{0}}^{1}=\braket{D^{0}1|\hat{\rho}|D^{0}1},\nonumber\\
&P_{D^{+-(-+)}}^{n}=\braket{D^{+-(-+)}n|\hat{\rho}|D^{+-(-+)}n},\quad n\geq 2,
\label{prob}
\end{align}
respectively. By further introducing vectors $\mathbf{P}_{X}=(P_{X}^{0}\hspace{3mm}P_{X}^{1}\hspace{3mm}P_{X}^{2}....)^{T}$,
with $X=00,S_{1},S_{2}$, containing the probabilities for states with one or both DQD:s
unoccupied and vectors $\mathbf{P}_{D}=(P_{D}^{2}\hspace{3mm}P_{D}^{3}
....)^{T}$ and $\quad\mathbf{P}_{D^{+-(-+)}}=(P^{2}_{D^{+-(-+)}}\hspace{3mm}P^{3}_{D^{+-(-+)}} ....)^{T}$
containing the probabilities for states with both DQD:s occupied, the ME
can be written
\begin{widetext}
\begin{align} 
\frac{d}{dt}\underbrace{\left(\begin{array}{lc}
\bold{P}_{00}\nonumber\\
\bold{P}_{S_{1}}\nonumber\\
\bold{P}_{S_{2}}\nonumber\\
P^{0}_{D}\nonumber\\
P^{1}_{D^{0}}\nonumber\\
P^{1}_{D}\nonumber\\
\bold{P}_{D}\nonumber\\
\bold{P}_{D^{+-}}\nonumber\\
\bold{P}_{D^{-+}}\nonumber\\
\end{array}\right)}_{\bold{P}}=\underbrace{\left(\begin{array}{ccccccccc}
M_{00} & M_{S_{1}}^{00}& M_{S_{2}}^{00}& 0 &0 &0 &0&0&0\nonumber\\
M^{S_{1}}_{00}& M_{S_{1}}& 0&  M^{S_{1}}_{D0}& M^{S_{1}}_{D^{0}1} &M^{S_{1}}_{D1}& M^{S_{1}}_{D} & M^{S_{1}}_{D^{+-}}& M^{S_{1}}_{D^{-+}} \nonumber\\
M_{00}^{S_{2}}& 0& M_{S_{2}}&  M^{S_{2}}_{D0}& M^{S_{2}}_{D^{0}1} &M^{S_{2}}_{D1}& M^{S_{2}}_{D} & M^{S_{2}}_{D^{+-}}& M^{S_{2}}_{D^{-+}} \nonumber\\
0& M_{S_{1}}^{D0}& M_{S_{2}}^{D0}& M_{D0} &0 &0 &0&0&0\nonumber\\
0& M_{S_{1}}^{D^{0}1}& M_{S_{2}}^{D^{0}1}& 0 &M_{D^{0}1} &0 &0&0&0\nonumber\\
0& M_{S_{1}}^{D}& M_{S_{2}}^{D1}& 0 &0 &M_{D1} &0&0&0\nonumber\\
0& M_{S_{1}}^{D}& M_{S_{2}}^{D}& 0 &0 &0 &M_{D}&0&0\nonumber\\
0& M_{S_{1}}^{D^{+-}}& M_{S_{2}}^{D^{+-}}& 0 &0 &0 &0&M_{D^{+-}}&0\nonumber\\
0& M_{S_{1}}^{D^{-+}}& M_{S_{2}}^{D^{-+}}& 0 &0 &0 &0&0&M_{D^{-+}}
\end{array}\right)}_{M}\left(\begin{array}{lc}
\bold{P}_{00}\nonumber\\
\bold{P}_{S_{1}}\nonumber\\
\bold{P}_{S_{2}}\nonumber\\
P^{0}_{D}\nonumber\\
P^{1}_{D^{0}}\nonumber\\
P^{1}_{D}\nonumber\\
\bold{P}_{D}\nonumber\\
\bold{P}_{D^{+-}}\nonumber\\
\bold{P}_{D^{-+}}\nonumber\\
\end{array}\right).\\
\label{ME}
\end{align}
The submatrices in $M$ above and below the diagonal are here given by 
\begin{align}
&(M_{00}^{S_{i}})_{nm}=\sum_{j=0,1}\delta_{nm+j}\mathcal{G}_{i}^{j},\hspace{2.5mm}(M_{S_{i}}^{Z})_{nm}=\sum_{j}\delta_{nm+j}x^{Y}_{nj}\mathcal{G}_{\bar{i}}^{j},\hspace{2.5mm}x^{D}_{nj}=\frac{4n+3-2j}{4(2n+1)},\hspace{2.5mm}x^{D^{+-}}_{nj}=\frac{n+j}{2n+1},\hspace{2.5mm}x^{D^{-+}}_{nj}=\frac{1}{4}\nonumber\\
&(M_{S_{i}}^{D0})_{1m}=\delta_{1m}\mathcal{G}_{i}^{0},\quad(M_{S_{i}}^{D^{0}1})_{1m}=\delta_{1m}\mathcal{G}_{i}^{1}/2,\quad(M_{S_{i}}^{D1})_{1m}=\delta_{1m}\mathcal{G}_{i}^{1}/4
(M_{D0}^{S_{i}})_{n1}=\delta_{n1}\tilde{\mathcal{G}}_{i}^{0}.
\end{align}
and 
\begin{align}
&(M_{S_{i}}^{00})_{nm}=\sum_{j=0,1}\frac{\delta_{nm-j}}{2-\delta_{n1}\delta_{j1}}\tilde{\mathcal{G}}_{i}^{j},\quad(M_{Z}^{S_{i}})_{nm}=\sum_{j=0,1}\delta_{nm-j}y^{Z}_{nj}\tilde{\mathcal{G}}_{\bar{i}}^{j},\nonumber\\
&y^{D}_{nj}=\frac{(4n+3+2j)}{4(2n+1+2j)},\hspace{2.5mm}y^{D^{+-}}_{nj}=\frac{n+j}{2n+1+2j},\hspace{2.5mm}y^{D^{-+}}_{nj}=\frac{1}{2},\nonumber\\
&(M_{D^{0}1}^{S_{i}})_{n1}=\delta_{n1}\tilde{\mathcal{G}}_{i}^{0}/2+\delta_{n2}\tilde{\mathcal{G}}_{i}^{1}/2,\quad
(M_{D1}^{S_{i}})_{n1}=\delta_{n1}\mathcal{G}_{i}^{0}/4+\delta_{n2}3\mathcal{G}_{i}^{1}/4,
\end{align}
respectively. Here $\mathcal{G}_{i}^{0}=\sum_{\nu}\bar{\Gamma}_{\nu
  gi}f_{\nu i}(-\hbar\omega/2)$, $\mathcal{G}_{i}^{1}=\sum_{\nu}\bar{\Gamma}_{\nu
  ei}f_{\nu i}(\hbar\omega/2)$, 
$\tilde{\mathcal{G}}_{i}^{0}=\sum_{\nu}\bar{\Gamma}_{\nu gi}\tilde{f}_{\nu
  i}(-\hbar\omega/2)$,
$\tilde{\mathcal{G}}_{i}^{1}=\sum_{\nu}\bar{\Gamma}_{\nu ei}\tilde{f}_{\nu
  i}(\hbar\omega/2)$ and we use the i-index convention $\bar{1}=2$ and $\bar{2}=1$. The submatrices on the
diagonal in $M$ are diagonal with elements such that the sum of every column in $M$ is zero. This structure ensures that the ME conserves probability.
\end{widetext}
We find the steady-state solution to the ME in Eq. \eqref{ME} analytically in three limiting cases:\\
$(i)$ For $\theta_{1}=\theta_{2}$,
symmetric bias voltages $eV_{1},eV_{2}\gg\hbar\omega$ applied across both DQD1 and DQD2.\\
$(ii)$ For the bias condition of Sec. \ref{nonLocal} with
$\Gamma_{1}\gg\Gamma_{2}$ and symmetric bias voltage across DQD1.
Here we calculate\ the distribution to zeroth order in $\Gamma_{2}/\Gamma_{1}$.\\
$(iii)$ For the bias condition of Sec. \ref{nonLocal} with
$\Gamma_{2}\gg\Gamma_{1}$ and symmetric bias voltage across DQD1.
The distribution is here calculated to first order in  $\Gamma_{1}/\Gamma_{2}$.\\ 
The solution for case $(i)$ is used to derive the stability condition in
Sec. \ref{derQME}, while the solutions for the cases $(ii)$ and $(iii)$ are used to obtain
the analytical expression for the current in DQD2, correct to first order in
$\Gamma_{2}/\Gamma_{1}$ and $\Gamma_{1}/\Gamma_{2}$, respectively. (See Sec.\ref{nonLocal}).

To find the solution of the ME in case $(i)$ we use a general property of
the ME. This property states that $P_{S_{1}}^{n}$ and $P_{S_{2}}^{n}$ will couple only to
the probabilities for states with both DQD:s unoccupied having $n$ or $n+1$ excitations
and for states with both DQD:s occupied having $n$ or $n-1$ excitations (See 
Fig. \ref{Tavisspec}). In turn the probabilities for these states will
couple to $P_{S_{1}}^{n+k}$ and $P_{S_{2}}^{n+k}$, with $k=-1,0,1$. Two coupled second order
difference equations are thus obtained for the probabilities in
$\mathbf{P}_{S_{1}}$ and $\mathbf{P}_{S_{2}}$. For the special conditions $(i)$
these difference equations become particularly simply. Introducing the vector
$\tilde{\mathbf{P}}_{n}=(P_{S_{1}}^{n}\hspace{3mm}P_{S_{2}}^{n})^{T}$ these equation be written
\begin{align}
&\sin^{4}(\theta)M_{0}\tilde{\bold{P}}_{1}=\left(2\cos^{4}(\theta)M_{0}+[1+\cos^{2}(\theta)]B\right)\tilde{\bold{P}}_{0}\nonumber\\
&\sin^{4}(\theta)M_{1}\tilde{\bold{P}}_{2}=\left[\cos^{4}(\theta)M_{1}+\cos^{4}(\theta)M_{0}+\frac{3B}
{2}\right]\tilde{\bold{P}}_{1}\nonumber\\
&-2\cos^{4}(\theta)M_{0}\tilde{\bold{P}}_{0},\nonumber\\
&\sin^{4}(\theta)M_{n}\tilde{\bold{P}}_{n+1}=\left[\cos^{4}(\theta)M_{n}+\sin^{4}(\theta)M_{n-1}+\frac{3B}
{2}\right]\tilde{\bold{P}}_{n}\nonumber\\
&-\cos^{4}(\theta)M_{n-1}\tilde{\bold{P}}_{n-1},\quad n\geq 2
\label{difference}
\end{align}
with 
\begin{align}
&B=\Gamma_{1}\Gamma_{2}\left(\begin{array}{cc} 
1 &-1 \\
 -1 & 1\end{array}\right),\nonumber\\
&M_{n}=\left(\begin{array}{cc} 
\Gamma_{1}^{2}/2+\Gamma_{2}^{2}b_{n} & \Gamma_{1}\Gamma_{2}(1/2+b_{n})\\
 \Gamma_{1}\Gamma_{2}(1/2+b_{n}) &
 \Gamma_{2}^{2}/2+\Gamma_{1}^{2}b_{n}\end{array}\right),
\end{align}
and
\begin{align}
&b_{n}=\frac{(4n+3)(4n+1)}{4(2n+1)[4n+1+2\cos^{2}(\theta)]}+\frac{1}{4}\nonumber\\
&+\frac{n(n+1)}{2(2n+1)[n+\sin^{2}(\theta)]}.
\end{align}
Together with the condition $\tilde{\mathbf{P}}_{n}\rightarrow 0$ for
$n\rightarrow\infty$ Eq. \eqref{difference} has the solution
$\tilde{\mathbf{P}}_{n}=\cot^{4n+2}(\theta)(1\hspace{3mm}1)^{T}P_{11}^{0}$,
independent on $\Gamma_{1}$ and $\Gamma_{2}$. This solution is then
used to find the probabilities for the states with both DQD:s occupied and
unoccupied, respectively. Requiring that the solution to Eq. \eqref{ME} is normalized we get
\begin{align}
&P_{00}^{n}=\cot^{4(n+1)}(\theta)x_{0},\quad P_{S_{i}}^{0}=\cot^{2}(\theta)x_{0},\nonumber\\
&P_{S_{i}}^{n}=2\cot^{4n+2}(\theta)x_{0},\quad n\geq 1\quad
&\nonumber\\
&P_{D}^{0}=x_{0},\quad P_{D}^{1}=2P_{D^{0}}^{1}=2\cot^{4}(\theta)x_{0},\nonumber\\ 
&P_{D}^{n}=2P_{D^{+-(-+)}}^{n}=2\cot^{4n}(\theta)x_{0},\quad n\geq 2,
\label{MEsymsolution}
\end{align}
with $x_{0}=[1-\cot^{4}(\theta)]/[1+\cot^{2}(\theta)+\cot^{4}(\theta)]^{2}$. It
is clear from Eq. \eqref{MEsymsolution} that a well defined solution, or
equivalently a solution with non-infinite mean number of excitations, exists if
 $\theta>\pi/4$. On physical grounds we argue that there exists a more general
 stability condition applying also for mixing angles
 $\theta_{1}\neq\theta_{2}$. We start by considering the case when
 $\theta_{1}=\theta_{2}>\pi/4$ and one of the mixing angles is increased. From the discussion below
Eq. \eqref{QME} it is clear that this will increase absorption relative to emission of photons by tunneling
 electrons. The mean number of excitations will then be decreased and the
 distribution must therfore still be convergent. Since all pairs of mixing
 angles $\theta_{1},\theta_{2}>\pi/4$ can be reached this way it follows that
 a well defined solution
 exists for all of them. The mean number of excitations will
 also decrease if the bias voltage is decreased. The stability condition,
 $\theta_{1},\theta_{2}>\pi/4$, must therefore also hold for finite bias
 voltages. This conclusion is further supported by numerical investigations.

In the limiting case $(ii)$ the relation between probabilities for states
with DQD2 unoccupied, e.g. $P_{00}^{n}$ and $P_{S1}^{m}$, are entirely
determined by the tunneling in DQD1.  Similarly he tunneling in DQD1 entirely determines the the relation between probabilities for states with DQD2
occupied. The tunneling in DQD2 only effects the total probability for DQD2 being occupied. The main steps in
the solution of the ME in this limit are most clearly visualized by rewriting $M\mathbf{P}=0$ as
\begin{align}
\underbrace{\left(\begin{array}{cc}
\bar{M}_{00}^{(0)}+\Gamma_{2}/\Gamma_{1}\bar{M}_{00}^{(1)} & \Gamma_{2}/\Gamma_{1}\bar{M}_{10}^{(1)}\\
\Gamma_{2}/\Gamma_{1}\bar{M}_{01}^{(1)} & \bar{M}_{11}^{(0)}+\Gamma_{2}/\Gamma_{1}\bar{M}_{11}^{(1)}\end{array}\right)}_{\bar{M}}\underbrace{\left(\begin{array}{c}
\bar{\bold{P}}_{20}\\
\bar{\bold{P}}_{21}\end{array}\right)}_{\bar{\bold{P}}}=0,
\label{MEassym1}
\end{align}
where the vectors $\bar{\mathbf{P}}_{20}$ and $\bar{\mathbf{P}}_{21}$ contain the probabilities
for states with DQD2 unoccupied and occupied, respectively. The matrix $\tilde{M}$
further contains the transition rates in $M$ divided by $\Gamma_{1}$. 
The starting point of the derivation is to note that for $\Gamma_{2}=0$ the matrix
is block-diagonal, with blocks $\bar{M}_{ii}^{(0)}$, and that
$\det(\bar{M}_{ii}^{(0)})=0$. It then follows that the eigenvalue zero of
the matrix $\bar{M}$ is doubly degenerate for $\Gamma_{2}=0$. To find the
solution $\bar{\mathbf{P}}^{(0)}$ to Eq. \eqref{MEassym1} to zeroth order in
$\Gamma_{2}/\Gamma_{1}$, i.e. the limit of $\bar{\mathbf{P}}$ as $\Gamma_{2}/\Gamma_{1}\rightarrow 0$,
we must therefore apply degenerate perturbation theory generalized to ME matrices. The first step in
this procedure is to find the two linearly independent solutions to Eq. \eqref{MEassym1}, i.e. the eigenvectors
corresponding to the eigenvalue zero of $\bar{M}$. Setting $\Gamma_{2}=0$ in
the equation the two linearly independent eigenvectors acquire the forms $\bar{\mathbf{P}}_{0}=(\bar{\mathbf{P}}^{(0)}_{20}\hspace{3mm}0)^{T}$ and
$\bar{\mathbf{P}}_{1}=(0\hspace{3mm}\bar{\mathbf{P}}^{(0)}_{21})^{T}$, where
$\bar{\mathbf{P}}^{(0)}_{20}$ and $\bar{\mathbf{P}}^{(0)}_{21}$ fulfill the
equations $M_{00}^{(0)}\bar{\mathbf{P}}^{(0)}_{20}=0$ and
$M_{11}^{(0)}\bar{\mathbf{P}}^{(0)}_{21}=0$, respectively. Importantly these
equations can readily be solved analytically as they give difference equations
similar to Eq. \eqref{difference}. The next step is to express $\bar{\mathbf{P}}^{(0)}$ as a normalized linear combination of
these vectors, i.e. $c_{0}\bar{\mathbf{P}}_{0}+c_{1}\bar{\mathbf{P}}_{1}$,
to which $\bar{\mathbf{P}}$ tend as $\Gamma_{2}/\Gamma_{1}\rightarrow 0$,
. To do this we define the projectors $\mathcal{P}=
\bar{\mathbf{P}}_{0}(U_{0}\hspace{3mm}0)+\bar{\mathbf{P}}_{1}(0\hspace{3mm}U_{1})$
and $\mathcal{Q}=1-\mathcal{P}$. Here $(U_{0}\hspace{3mm}0)$ and
$(0\hspace{3mm}U_{1})$, with $U_{i}=(1\hspace{3mm}1\hspace{3mm}...)$, are the left eigenvectors to the eigenvalue zero in $\tilde{M}$
normalized so that $U_{i}\bar{\mathbf{P}}_{2i}=1$. Then by applying the steps presented in
Ref. \cite{sakurai} the equation
$\mathcal{P}M^{(1)}\mathcal{P}\bar{\mathbf{P}}^{(0)}=0$ for
$\bar{\mathbf{P}}^{(0)}$ is obtained. In turn this equation gives  
\begin{align}
&c_{0}U_{0}\bar{M}_{00}\bar{\bold{P}}_{20}+c_{1}U_{1}\bar{M}_{01}\bar{\bold{P}}_{21}=0,
\end{align}
which together with the normalization condition $c_{0}+c_{1}=1$,
determines $c_{0}$ and $c_{1}$. The different element in
$\bar{\mathbf{P}}^{(0)}$ can then be written 
\begin{align}
&P_{00}^{n}=[\gamma/(1-\gamma)]^{2n+1}\tilde{c}_{0}\nonumber\\
&P_{S_{1}}^{n}=(2-\delta_{n0})[\gamma/(1-\gamma)]^{2n}\tilde{c}_{0})\nonumber\\
&P_{S_{2}}^{n}=(2-\delta_{n0})[\gamma/(1-\gamma)]^{2n+1}\tilde{c}_{1}\nonumber\\
&P_{D}^{0}=\tilde{c}_{1},\quad P_{D}^{1}=2P_{D^{0}}^{01}=2[\gamma/(1-\gamma)]^{2}\tilde{c}_{1},\nonumber\\
& P_{D}^{n}=2P^{n}_{D^{+-(-+)}}=2[\gamma/(1-\gamma)]^{2n}\tilde{c}_{1},\quad n\geq 2,
\label{sol1}
\end{align}
where
\begin{align}
&\tilde{c}_{0}=\frac{\eta\gamma^{2}}{(1-2\gamma+2\gamma^{2})[1+2\cot^{2}(\theta_{1})]+\gamma^{2}},\nonumber\\
&\tilde{c}_{1}=\frac{\eta (1-\gamma^{2})[1+2\cot^{2}(\theta_{1})]}{(1-\gamma+\gamma^{2})[1+2\cot^{2}(\theta_{1})]+\gamma^{2}},
\end{align}
and $\eta=(1-2\gamma)/(1-\gamma+\gamma^{2})$. With the probabilities in Eq. \eqref{sol1} and the rates between
DQD2 and its right lead for the corresponding states it is
straightforward to derive the second expression for the current through DQD2 in Eq. \eqref{nonlocal}.

To solve the ME in case $(iii)$ we first recall, from the main text, that only the
state and transitions depicted in Fig. \ref{scheme} contribute to the
transport quantities to first order in $\Gamma_{1}/\Gamma_{2}$. To find the probability 
distribution to $\mathbf{P}$ first order in $\Gamma_{1}/\Gamma_{2}$ only the
additional transitions $\{\ket{D^{\alpha 1}}\}\rightarrow\ket{S_{2}0}$ need to
be taken into account.  Thus, the distribution is obtained by solving an
effective ME $M_{e}\mathbf{P}_{e}=0$. We get
\begin{align}
&P_{D}^{0}=\frac{1}{1+\gamma}-\frac{\Gamma_{1}}{\Gamma_{2}}\frac{2[2+\cos^{2}(\theta_{1})]}{\sin^{2}(\theta_{1})}\frac{\gamma^{2}}{(1+\gamma)^{2}}\nonumber\\
&+\frac{\Gamma_{1}}{\Gamma_{2}}\frac{[1+\cos^{2}(\theta_{1})]}{\sin^{2}(\theta_{1})}\frac{\gamma^{2}(1-\gamma)}{(1+\gamma)^{2}}\nonumber\\
&P_{S_{2}}^{0}=\frac{\gamma}{1+\gamma}-\frac{\Gamma_{1}}{\Gamma_{2}}\frac{2[2+\cos^{2}(\theta_{1})]}{\sin^{2}(\theta_{1})}\frac{\gamma^{3}}{(1+\gamma)^{2}}\nonumber\\
&-\frac{\Gamma_{1}}{\Gamma_{2}}\frac{[1+\cos^{2}(\theta_{1})]}{\sin^{2}(\theta_{1})}\frac{\gamma^{2}(1-\gamma)}{(1+\gamma)^{2}}\nonumber\\
&P_{D}^{1}=2P_{D^{0}}^{1}=\frac{2\Gamma_{1}}{\Gamma_{2}}\frac{1+\cos^{2}(\theta_{1})}{\sin^{2}(\theta_{1})}\frac{\gamma^{2}}{(1+\gamma)}\nonumber\\
&P_{S_{1}}^{0}=\frac{\Gamma_{1}}{\Gamma_{2}}\frac{\gamma^{2}}{(1+\gamma)}.
\label{sol2}
\end{align}
These probabilities are used to obtain the first expression for current
through DQD2 in Eq. \eqref{nonlocal}. 

\section*{Appendix D}
In this Appendix we present the derivation of currents, current correlations as
well as the full statistics of charge transfer accross the DQD:s. Following the
procedure of Refs. \cite{Flindt05,kiesslich06} we rewrite the QME of Eq. \eqref{QME} in the
n-resolved form and Fourier transform it with respect to the number of
electrons having tunneled through DQD1 and DQD2. The QME then transforms to $d\hat{\rho}/dt=\mathcal{L}(\chi_{1},\chi_{2})\hat{\rho}$,
where the counting fields $\chi_{1}$ and $\chi_{2}$ are the conjugate variables to the number of electrons having tunneled through DQD1 and DQD2. The eigenvalue of $\mathcal{L}(\chi_{1},\chi_{2})$ tending to
zero as $\chi_{1},\chi_{2}\rightarrow 0$ is the long time limit cumulant generating function
$\mathcal{F}(\chi_{1},\chi_{2})$. The currents and the noise
are obtained from the first and second derivatives of
$\mathcal{F}(\chi_{1},\chi_{2})$,
i.e. $I_{i}=e\partial_{i\chi_{i}}\mathcal{F}|_{\chi_{1}=\chi_{2}=0}$ and
$S_{ij}=e^{2}\partial_{i\chi_{i}}\partial_{i\chi_{j}}\mathcal{F}|_{\chi_{1}=\chi_{2}=0}$. These
quantities can conveniently be accessed via the eigenvalue problem
$\mathcal{L}(\chi_{1},\chi_{2})[\hat{\rho}(\chi_{1},\chi_{2})]=\mathcal{F}(\chi_{1},\chi_{2})\hat{\rho}(\chi_{1},\chi_{2})$. In
the present paper this full QME-approach is used only to calculate the current in Fig. \ref{fintemp}.

In the ME limit the eigenvalue problem above reads $M(\chi_{1},\chi_{2})\mathbf{P}(\chi_{1},\chi_{2})=\mathcal{F}(\chi_{1},\chi_{2})\mathbf{P}(\chi_{1},\chi_{2})$.
We use this equation to calculate the noise plotted in
Fig. \ref{resultnoise}. We also use the equation to obtain the CGF of Eq. \eqref{CGF}, i.e. the CGF to
first order in $\Gamma_{1}/\Gamma_{2}$ for the bias condition of
DQD2 described in Sec. \ref{nonLocal} with DQD1 in the high-bias regime. To do this we write $\det[M(\chi_{1},\chi_{2})-\mathcal{F}(\chi_{1},\chi_{2})]=0$ as 
\begin{align}
\det\left(\begin{array}{cc}
M_{e}(\chi_{1},\chi_{2})-\mathcal{F} & M_{r\rightarrow e}(\chi_{1},\chi_{2})\\
M_{e\rightarrow r}(\chi_{1},\chi_{2}) & M_{r}(\chi_{1},\chi_{2})-\mathcal{F}
\label{char1}
\end{array}\right)=0.
\end{align}
Here $M_{e\rightarrow r}$, $M_{r\rightarrow e}$ 
and $M_{r}$ are matrices describing the transitions to, from and within the
subspace of states not included in the Fig. \ref{scheme}. Importantly,
$(M_{e\rightarrow
  r})_{nm}\propto\mathcal{O}(\Gamma_{1}/\Gamma_{2})$ and $(M_{r\rightarrow
  e})_{nm}\propto\mathcal{O}(1)$. This means that
\begin{equation}
\det[M_{e}(\chi_{1},\chi_{2})-\mathcal{F}^{(1)}(\chi_{1},\chi_{2})]=0,
\label{char2}
\end{equation}
 where $\mathcal{F}^{(1)}(\chi_{1},\chi_{2})$ is the CGF to first order in
 $\Gamma_{1}/\Gamma_{2}$. Dropping terms of $\mathcal{O}[(\Gamma_{1}/\Gamma_{2})^{2}]$ in
Eq. \eqref{char2} and solving for $\mathcal{F}^{(1)}$ then gives Eq. \eqref{CGF}.

\end{document}